# Role of silicon and carbon on the structural and electrochemical properties of Si-$Ni_{3.4}Sn_4$-Al-C anodes for Li-ion batteries

Tahar Azib[a*+], Claire Thaury[a,b*], Cécile Fariaut-Georges[a], Thierry Hézèque[b], Fermin Cuevas[a], Christian Jordy[b] and Michel Latroche[a]

[a]Univ. Paris Est Creteil, CNRS, ICMPE, UMR7182, F-94320, Thiais, France.

[b]SAFT Batteries, 113 Bd. Alfred Daney, 33074 Bordeaux, France.

## Abstract

Varying the amounts of silicon and carbon, different composites have been prepared by ball milling of Si, $Ni_{3.4}Sn_4$, Al and C. Silicon and carbon contents are varied from 10 to 30 wt.% Si, and 0 to 20 wt.% C. The microstructural and electrochemical properties of the composites have been investigated by X-Ray Diffraction (XRD), Scanning Electron Microscopy (SEM) and electrochemical galvanostatic cycling up to 1000 cycles. Impact of silicon and carbon contents on the phase occurrence, electrochemical capacity and cycle-life are compared and discussed. For C-content comprised between 9 and 13 wt.% and Si-content ≥ 20 wt.%, Si nanoparticles are embedded in a $Ni_{3.4}Sn_4$-Al-C matrix which is chemically homogeneous at the micrometric scale. For other carbon contents and low Si-amount (10 wt.%), no homogeneous matrix is formed around Si nanoparticles. When homogenous matrix is formed, both $Ni_3Sn_4$ and Si participate to the reversible lithiation mechanism, whereas no reaction between $Ni_3Sn_4$ and Li is observed for no homogenous matrix. Moreover, best cycle-life performances are obtained when Si nanoparticles are embedded in a homogenous matrix. Composites with carbon in the 9-13 wt.% range and 20 wt.% silicon lead to the best balance between capacity and life duration upon cycling. This work experimentally demonstrates that embedding Si in an intermetallic/carbon matrix allows to efficiently accommodate Si volume changes on cycling to ensure long cycle-life.

*Both authors contributed equally to this manuscript

+ Corresponding author

• E-mail address: azib@icmpe.cnrs.fr

• Full postal address: CNRS, ICMPE, UMR7182, 2 rue Henri Dunant, F-94320, Thiais, France

## 1. Introduction

Energy storage demands for current and future portable electronic devices and electrical vehicles are ever growing. Performances of Li-ion batteries in terms of specific power and density surpass other rechargeable battery systems but are still unsatisfactory for many applications such as extended-range electric vehicles [1–3]. The development of advanced anode materials exhibiting higher capacity than current graphite electrodes ($C$ = 370 mAhg$^{-1}$) remains very challenging [4]. Presently a lot of materials are studied as alternatives for carbonaceous ones to enhance the energy density of the negative electrode using various strategies such as conversion or alloying reactions [5,6]. For the latter, a special interest is brought to *p*-bloc elements that form lithiated compounds [7]. For instance Sn and Si can form $Li_{4.4}Sn$ and $Li_{3.75}Si$ compounds to store respectively 994 and 3600 mAhg$^{-1}$ in electrochemical units, leading to much higher capacities than that of graphite [8,9]. In addition to their high theoretical capacity, these elements exhibit low potential and environmental friendliness, both suitable as anode properties.

However, electrodes made of pure *p*-bloc elements are considered inappropriate because of large capacity fading during electrochemical cycling [10]. The large volume expansion that accompanies Li insertion and extraction during cycle life results in electrode pulverization and loss of electrical contact [11,12]. To ensure mechanical stability, embedding capacitive elements into a buffering matrix that helps accommodating volume changes, while improving electronic conduction, has been proposed [13,14]. This concept can be implemented, for instance, using binary intermetallic compounds made of one element that reacts with lithium (*e.g.* Si or Sn) and another one that is inactive (*e.g.* Fe, Ni, Co or Cu). Some reported intermetallics to achieve this goal are $Ni_3Sn_4$ [15–17], $Cu_6Sn_5$ [18], $CoSn_2$ [19,20], $FeSn_2$ [20], $FeSi_2$ [21] and $NiSi_2$ [22]. This approach has been successfully extended to ternary silicides such as $Ti_4Ni_4Si_7$ [23,24], $TiFeSi_2$ [25] and $Ti_3Si_2C_2$ [26]. Moreover, binary and ternary intermetallics have been associated to carbonaceous materials to form highly efficient multi-component systems, such as Si/$FeSi_2$/C [27] and Si/$Cu_3Si$/C [28], for further improvement of both the mechanical cohesion of silicon and the electronic conductivity of the whole active material.

In this context, an efficient composite Si-$Ni_{3.4}Sn_4$-Al-C has been already synthetized by mechanochemistry [29]. It consists of sub-micrometric Si particles embedded in a nanostructured and complex matrix mainly formed by $Ni_{3.4}Sn_4$ intermetallic compound and disordered carbon. Si and Sn are the capacitive elements whereas Ni and C are expected to accommodate the volume changes and to insure good electronic as well as ionic conductivity of the electrode material. Al addition is based on reported beneficial effects on the cycle-life of Si-Sn-Al amorphous electrodes [30], best results being obtained for the composition 30Si-67Sn-3Al which contains 3 wt.% of Al. This composite exhibits high reversible capacity (~ 700 $mAhg^{-1}$) and long cycle life (over 280 cycles) [31,32]. Here we investigate the influence of silicon and carbon contents on the microstructure of this composite material and its electrochemical properties. Silicon and carbon contents were varied from 10 to 30 wt.% Si, and from 0 to 20 wt.% C, respectively.

## 2. Experimental

Seven nanostructured composites Si-$Ni_{3.4}Sn_4$-Al-C with various carbon and silicon contents have been synthetized. Their overall compositions (weight percent) are given in Table 1. Two series were considered: i) for constant Si-content (~20 wt.%), the carbon amount was increased from 0 to 20 wt.% and ii) for constant C-content (13 wt.%), the silicon content was increased from 10 to 30 wt.%. Variation of C and Si contents were counterbalanced by the amount of $Ni_{3.4}Sn_4$ intermetallic compound.

The composites were made in two steps. Firstly, a tin-based intermetallic precursor $Ni_{3.4}Sn_4$ was synthesized by powder metallurgy followed by mechanical milling of this intermetallic with Si, Al and C elements. To synthesize $Ni_{3.4}Sn_4$, elemental nickel (99.9 %, <45 µm, Cerac) and tin (99.9 %, <45 µm, Alfa-Aesar) powders were mixed in 3.4:4 atomic ratio, pelleted and sintered under argon atmosphere in a silica tube at 700°C for 7 days. This annealing time was necessary to obtain single-phase compound. The pellets were mechanically pulverized down to 125 µm size. Then, powders of silicon (99.9 %, <1 µm), aluminum (99 %, ≤ 75 µm, Aldrich), graphite and $Ni_{3.4}Sn_4$ were weighted according to the different compositions given in Table 1 and mechanically milled in a Fritsch Pulverisette 7 planetary mill. Ball milling was performed under argon atmosphere for 20 hours in an 80 ml volume jar with 7 mm diameter balls. The ball-to-powder weight ratio was 5:1. Both jar and balls are made of stainless steel. Vial rotation speed of the mill was 600 rpm.

The crystal structure of the milled composites was analyzed by XRD with a Bruker D8 advance θ-θ diffractometer using Cu Kα radiation. Diffraction patterns were analyzed by the Rietveld method using the FullProf software [33]. Phase assignation and refinement of their crystal structure including lattice parameters, phase amounts, peak-profile factors were undertaken. Crystal sizes were evaluated from the Lorentzian contribution to the Thompson–Cox–Hastings pseudo–Voigt profile-shape function used in the Rietveld analysis, after considering the instrumental contribution [34]. For $Ni_{3.4}Sn_4$, variations of the Ni over-stoichiometry were obtained from the refinement of partial occupancy of Ni at site 2*c* (0, 0, ½) in *C*2/*m* space group [35] . The microstructure of 20h-milled composites (*i.e.* phase distribution and morphology) was studied by scanning electron microscopy (SEM) using a SEM-FEG MERLIN ZEISS device. The SEM images were collected in Back-Scattered Electron (BSE) mode at cross-sections of the ball-milled composite powders. The composites were embedded in epoxy resin then polished and metalized with platinum. Point-by-point composition analysis was carried out by Energy-Dispersive-X-ray (EDX) analysis.

Electrochemical measurements were carried out by galvanostatic cycling in coin-type half-cells. Working electrode material was prepared by mixing 55 wt.% of the composite sieved under 125 µm, 20 wt.% of carboxymethyl-cellulose (CMC) binder and 25 wt.% of carbon black. Low loading of the composite material (55 wt.%) was selected to avoid limitations on its intrinsic electrochemical properties due to electrode formulation. Metallic lithium was used as counter negative electrode. The electrolyte consisted of 1 M solution of $LiPF_6$ dissolved in ethylene carbonate (EC)/propylene carbonate (PC)/dimethyl carbonate (DMC) (1:1:3 vol./vol./vol.). A polyolefin Celgard® microporous membrane and a nonwoven polyolefin membrane were used as separators. The battery was assembled in an argon-filled glove box. The experiments were performed using a Biologic VMP3 potentiostat at a typical cycling-rate of C/10 (full capacity C in 10 hours) with potential window comprised between 70 mV and 2 V. Different conditions were used for the first three cycles and every 20 cycles. To fully activate Si particles, the first cycle was performed at slow rate C/50 with a cut-off voltage of 0 V. The second cycle was performed at medium rate C/20 with a cut-off voltage of 10 mV. This cycle aimed to get to obtain the maximal capacity avoiding lithium deposition on the active material. The third cycle was also done at C/20 but with higher cut-off potential (70 mV) to improve cycle-life. Cycles performed following the latter protocol are named hereafter reference cycles. They were done on the third and every twenty cycles. Otherwise specified, only reference cycles are reported in this paper.

## 3. Results

### 3.1. Structural and microstructural characterization

All composites have been characterized by X-ray diffraction and Table 2 gathers the structural and microstructural data obtained from the Rietveld analysis. Graphical output of the Rietveld analysis for each composite with their corresponding Rietveld agreement factors are gathered in Fig. S1 (Supplementary Information).

A carbon-free composite was first synthetized with the nominal composition $Si_{0.22}Ni_{0.22}Sn_{0.53}Al_{0.03}$. XRD patterns of this sample before and after 20-hour milling are displayed in Fig. 1. The observed diffraction lines can be attributed to different phases shown at the bottom of the figure. Before milling, main peaks are indexed with diffraction lines belonging to Si (S.G. = $Fd\bar{3}m$, $a$ = 5.430 Å) and $Ni_{3.4}Sn_4$ (S.G. = $C2/m$, $a$ = 12.37 Å, $b$ = 4.069 Å and $\beta$ = 104.06) [36]. After milling, besides minor contribution of initial reactants Si and $Ni_{3.4}Sn_4$, main diffraction peaks in the XRD pattern can be indexed with two novel phases: $NiSi_2$ (S.G. = $Fm\bar{3}m$, $a$ = 5.406 Å) and β-Sn (S.G. = $I4_1/amd$, $a$ = 5.832 Å, $a$ = 3.182 Å )[36]. The formation of both phases is attributed to the mechanochemically induced reaction $R_1$ between the intermetallic and the silicon as follows:

$$6.8\,Si + Ni_{3.4}Sn_4 \rightarrow 3.4\,NiSi_2 + 4\,Sn \qquad \text{R1}$$

As the diffraction peaks of Si and $NiSi_2$ strongly overlap, it is hard to detect $NiSi_2$ formation by XRD. The occurrence of β-Sn is easier to be detected by XRD due to its high scattering power, though main diffraction peaks of β-Sn partially overlap with those of $Ni_{3.4}Sn_4$. Thus, β-Sn diffraction peaks can serve as markers of reaction $R_1$ [31]. Note that no diffraction peaks related to Al have been detected. The low Al content used (3 wt.%), Al amorphization or dissolution into other detected phases may explain this result. This is a common feature of all composites synthetized in this research and will not be further discussed.

Fig. 2 shows the XRD patterns of the composites ranging from 9 to 20 wt.% of carbon for constant Si content (20 wt.% Si). After 20 hours of milling, diffraction lines from $Ni_{3.4}Sn_4$, β-Sn-phase and Si/$NiSi_2$ are identified. Changes with C-content in the relative intensity of diffraction peaks at 2θ ~44 indicate minor occurrence of reaction $R_1$, especially for low C-content. Indeed, Rietveld analysis results (Table 2) show that on increasing the C-content from 0 to 20 wt.%, the Sn and $NiSi_2$ amounts decrease from 42 to 0.7 wt.% and from 28 wt.% to no detection, respectively. This reveals that the presence of carbon significantly reduces the extent of reaction $R_1$.

Significant peak broadening of intermetallic $Ni_{3.4}Sn_4$ is also observed after milling. It is attributed to crystallite size reduction and this phenomenon is more significant for the lower carbon contents. The crystallite size *L* of $Ni_{3.4}Sn_4$ varies from 8 to 50 nm when C-content increases from 0 to 20 wt.%. In addition, the Ni over-stoichiometry *x* of the intermetallic phase decreases from 0.6 to 0.1 on reducing the C-content (Table 2), which concurs with the homogeneity domain of $Ni_{3+x}Sn_4$ [35]. These results evidence that the crystallographic properties of the intermetallic phase are strongly dependent on the use of carbon as process control agent (PCA) during milling. When the C-content decreases, the crystallite size of the intermetallic phase diminishes, and Ni atoms partially segregate out of $Ni_{3.4}Sn_4$.

Fig. 3 shows the XRD patterns of the composites ranging from 10 to 30 wt.% of silicon (*i.e.* at a constant 13 wt.% C-content). Whatever the Si content, peak broadening is very similar for the three compositions and Rietveld analyses lead to comparable crystallite sizes. After milling, diffraction lines from $Ni_{3.4}Sn_4$ and Si are still clearly identified. Nonetheless, as for the carbon series, changes in the Sn content related to reaction $R_1$ are observed. The Sn amount gradually increases from 0.5 wt.% to 6 wt.% when Si content increases from 10 to 30 wt.%.

Fig. 4 shows the cross-section SEM-BSE micrographs of the carbon-containing composites milled for 20 hours. At constant Si-amount (20 wt.%), from top to bottom of Fig. 4, the microstructure of low-content carbon composites (9 and 13 wt.%) consist of spherical dark areas, ~ 100 nm in size, surrounded by a grey matrix. In contrast, the richer carbon composites (17 and 20 wt%) show a mixture of dark, grey and micrometric bright areas of variable size. EDX analysis has been done for the 9 wt.% carbon-containing composite to identify the phases related to the different contrasts (Fig. 5). Dark and bright areas correspond to silicon and $Ni_{3.4}Sn_4$, respectively, whereas grey parts contain all elements: silicon, nickel, tin, aluminum and carbon, which are intimately mixed at the spatial resolution of EDX point-analysis (< 100 nm). Therefore, for low carbon contents (9 and 13 wt.%) Si nanoparticles are embedded in a homogenous and multi-elementary matrix, whereas at high carbon contents (17 and 20 wt.%) a heterogeneous mixture of Si nanoparticles, micrometric $Ni_{3.4}Sn_4$ and C-containing areas is observed.

At constant C-amount (13 wt.%), from left to right of Fig. 4, the 10 wt.% silicon composite consists in silicon (dark) nanoparticles surrounded by heterogeneous mixture of micrometric $Ni_{3.4}Sn_4$ (bright) and carbon-containing areas. For 20 and 30% of silicon, the microstructure of the composites is very similar to that of the lowest carbon content (9 wt.%) with silicon particles (dark) embedded in a homogeneous and multi-elementary matrix (grey).

As it will be later shown, the formation of the multi-elementary and homogenous matrix around Si nanoparticles is a key issue for the cycling stability of the composite. It should be noted that, based on XRD analysis (Fig. 2 and Fig. 3), Ni and Sn elements in this matrix are associated in the form of nanostructured $Ni_{3.4}Sn_4$ phase.

### 3.2. Electrochemical properties

The capacity retention on galvanostatic cycling has been initially analyzed for the five composites at constant Si-amount (20 wt.%) and increasing carbon content from 0 to 20 wt.%. Table 3 gathers relevant electrochemical properties in terms of reversible and irreversible capacities as well as capacity fading on cycling compared to the $C_{upper\ limit}$. This upper limit is determined from the overall composite composition (Table 1) and assuming that all active elements can fully react with lithium according to their individual electrochemical capacities:

3600, 994, 990 and 370 $mAhg^{-1}$ for Si, Sn, Al and C, respectively [7]. For the carbon-free composite, the evolution of reversible capacity on cycling for reference cycles is shown in Fig. S2 (Supplementary Information). Though it provides high discharge capacity in the first cycle, $C_{lith,1st}$ = 949 $mAhg^{-1}$, this composite suffers from severe capacity-loss on cycling with a final capacity of $C_{lith,206th}$ = 210 $mAhg^{-1}$. For 9 to 17 wt.% C composites, almost 100% of upper-limit capacity is recovered in the first lithiation (95, 98 and 97% for 9, 13 and 17wt.% C composites, respectively). Interestingly, all these 3 composites have low irreversibility, less than 18% in the first cycle. For C-containing composites, the evolution of reversible capacity upon cycling is displayed in Fig. 6. All composites have slightly lower capacities than the upper limit and high reversible capacities around 700 $mAhg^{-1}$ at the third cycle, except for the highest carbon content, 20 wt.%, that drops to 520 $mAhg^{-1}$ (Table 3). As concerns capacity fading upon cycling, it increases regularly from 0.024 up to 0.073 %/cycle with carbon content. Finally, coulombic efficiencies have been determined (Fig. S3, Supplementary Information). They are close to 100% without clear differences with the composition.

Fig. 7a shows the potential profiles of the first galvanostatic cycle for the 9 and 20 wt.%-carbon composites. For the first discharge, upon lithiation, capacities of 1196 and 923 $mAhg^{-1}$ are achieved for 9 and 20 wt.% of carbon, respectively. Upon charging, capacities of 985 and 702 $mAhg^{-1}$ are obtained. Two different potential features can be noticed in the discharging profiles. For the 9 wt.%-C composite, a sharp potential drop between $E_{ini}$ 1.5 V and 0.5 V is first observed. Then, a large pseudo-plateau starting from 0.5 V down to the end of discharge (0 V) occurs. This plateau is attributed to the gradual formation of Li-Sn and Li-Si alloys by lithiation of $Ni_{3.4}Sn_4$ and Si, respectively. In contrast, for the 20 wt.%-C composite, the potential decrease quickly from 1.5 V to 0.15 V, followed by a shorter sloping plateau down to 0 V. The polarization is larger in the richest carbon composite. The irreversible capacities attributed to the formation of the SEI layer at the first cycle are 18 and 24% for 9 and 20 wt.% C composites, respectively.

Fig. 7b shows the dependence of differential capacity (*dQ*/*dV*) *vs* potential for the two-representative carbon-containing composites (9 and 20 wt.%) at the third cycle. For the carbon-rich one (20 wt.%), two main peaks are observed both during lithiation and delithiation. The lithiation peaks at 0.24 and 0.09 V are characteristic of the alloying reaction of lithium with amorphous silicon to form $Li_ySi$ amorphous phases. Amorphous silicon is widely reported to be formed during the first cycle [37,38]. Jimenez *et al.* [39] reported that the reduction peak at 0.24 V is associated with the formation of LiSi from the amorphous Si and subsequent phase transition from LiSi to $Li_7Si_3$. The second reduction peak at 0.09 V is attributed to the formation of $Li_{13}Si_4$ phase. For the delithiation sweep, two distinct peaks appear at 0.28 V and 0.49 V. They are attributed to the full decomposition of $Li_ySi$ phases into silicon [40,41]. It should be noted that no evidence of lithiation/delithiation peaks due to the reaction between $Ni_{3.4}Sn_4$ and lithium are detected for the 20 wt.%-C composite. For the carbon poor composite (9 wt.%-C), broad lithiation/delithiation peaks are observed for Si at 0.18/0.46 V and 0.09/0.28 V for reduction/oxidation reactions, respectively. In addition, extra peaks at 0.3 V for lithiation and 0.57 V for delithiation are observed. This redox pair is attributed to $Li_7Sn_2$ formation/decomposition resulting from the conversion reaction between intermetallic $Ni_{3.4}Sn_4$ and Li [17].

Regarding composites with constant C-amount (13wt.%) and increasing Si-content from 10 to 30 wt.%, the evolution of reversible capacity upon cycling is displayed in Fig. 8. Composites with 20 and 30 wt.% of Si exhibit reversible capacities close to 700 $mAhg^{-1}$ at the third cycle whereas 10%-Si gives much lower capacity (388 $mAhg^{-1}$) for the same cycle. The capacity fading over 400 cycles is 0.077 and 0.082 %/cycle for 10 and 30 wt.% Si, respectively but only 0.039 %/cycle for 20 wt.% Si. Coulombic efficiencies have also been determined for the Si-series (Fig. S4, Supplementary Information). As for the C-series, they are close to 100% without clear differences with the composition.

Fig. 9a shows the voltage profiles at the first galvanostatic cycle for the 10 and 30 wt.%-Si composites. For the first discharge, capacities of 794 $mAhg^{-1}$ and 1186 $mAhg^{-1}$ are achieved. Upon charge 572 $mAhg^{-1}$ and 966 $mAhg^{-1}$ are recovered, respectively. On discharge, the potential gradually drops down to 0.5 V for both composites followed by a pseudo-plateau down to 0 V, which is larger for the Si-rich composite. The irreversible capacity for the Si- poor composite (28%) is higher than that of the Si-rich one (17%). Fig. 9b shows the *dQ/dV* dependence *vs* potential for the two Si composites (10 and 30 wt.%) at the third cycle. The two redox pairs related to Si lithiation/delithiation at 0.22/0.49V and 0.09/0.34 V as well as that related to the conversion reaction of $Ni_{3.4}Sn_4$ at 0.3/0.62 V are observed. They concur with the three redox pairs observed in the carbon-series (Fig.7b) leading to the formation/decomposition of LiSi-$Li_7Si_3$, $Li_{13}Si_4$ and $Li_7Sn_2$. It is worth to note that for the composite with 30wt.%, in addition to 0.3/0.62 V redox pair, new peaks appear at higher potentials that are attributed to the formation of Li-poor $Li_ySn$ alloys such as $Li_5Sn_2$, $Li_7Sn_3$ and LiSn [17,42]. The peak intensity is significantly lower for the composite with lowest Si-content (10 wt.%) as concerns not only $Li_ySi$ both also $Li_ySn$ signals. The latter result implies that reversible lithiation of $Ni_{3.4}Sn_4$ intermetallic, which is the major phase for the Si-poor composite (Table 2), is very limited.

## 4. Discussion

Seven composites synthetized by mechanochemistry of four components (Si, $Ni_{3.4}Sn_4$, Al and C) have been prepared with different amounts of silicon (from 10 to 30 wt.%) and carbon (from 0 to 20 wt.%). After 20 hours of milling, all composites consist of a mixture of ~100 nm in size nanostructured Si particles surrounded by a matrix that is composed of several phases: $Ni_{3.4}Sn_4$, C, $NiSi_2$ and Sn. The phase amount and crystallinity of the constituents of this matrix as well as its morphology depend both on C and Si-contents. As shown in Fig. 4, for high C-content (17 and 20 wt.%) and low Si-content (10 wt.%) the matrix is chemically heterogeneous at the spatial resolution of SEM-BSE analysis, showing large micrometric $Ni_{3.4}Sn_4$ areas. In contrast, for low C-content (9 and 13 wt.%) and high Si-content (20 and 30 wt.%), all matrix constituents are intimately mixed at the nanoscale and form a homogeneous matrix, mainly composed of $Ni_{3.4}Sn_4$ and disordered carbon, in which the Si nanoparticles are embedded. In the latter nanocomposites, crystallite sizes for Si and $Ni_{3.4}Sn_4$ remain in the low nanocrystalline range with $L_{Si} = 12 \pm 2$ nm and $L_{Ni_{3.4}Sn_4} = 8 \pm 1$ nm (Table 2).

Despite compositional and microstructural differences, the electrochemical voltage profiles are alike for all C-containing composites as regards to the reaction between Si and lithium (Fig. 7 & Fig. 9). In contrast, reversible lithiation of $Ni_{3.4}Sn_4$ is only detected when the intermetallic domains are small, *i.e.* when a homogenous matrix is formed. This concurs with previous studies on $Ni_{3.4}Sn_4$ nanostructuration [17], showing that micrometric $Ni_{3.4}Sn_4$ exhibits very limited reactivity with Li providing low reversible capacity (below 25 $mAhg^{-1}$), while nanometric one is much more reactive, reaching 240 $mAhg^{-1}$ for a crystallite size of 6 nm, comparable with present result (Table 2). Reactivity of both Si and $Ni_{3.4}Sn_4$ phases allows approaching the first lithiation capacity to the upper-limit capacity with ratio exceeding 95% (Table 3). The $dQ/dV$ dependence *vs* potential curves confirms the reactivity of Si and $Ni_{3.4}Sn_4$ phases [43,44], being the latter only observed for low C contents and rich Si ones. Interestingly, when the intermetallic phase reacts (*i.e.* for homogenous matrix), the Si lithiation/delithiation peaks are broader (9 wt.% C in Fig. 7b). This can be tentatively attributed to the confinement effect of Si-lithiation by the surrounding matrix.

The present results clearly demonstrate that the Si/C ratio influences both microstructural and electrochemical properties. First, the role plaid by carbon for the phase stability and structuration of the composites at the milling stage is crucial. For the carbon-free composite, silicon and $Ni_{3.4}Sn_4$ react according to $R_1$ and form tin and binary phase $NiSi_2$. This reaction is detrimental for efficient composite nanostructuration on milling due to the high ductility of tin. Indeed, relatively large crystallite size is observed for Sn ($L_{Sn}$ = 27 nm) in C-free composite. Moreover, as concerns electrochemical properties, $NiSi_2$ is reported to be inactive toward lithium [22] and tin may suffer from agglomeration effects on cycling [45]. These two latter properties are at the origin of the poor electrochemical cycling performance of C-free composite. These issues were solved by carbon addition: with 9 wt.% of carbon, Sn formation decreases from 42 wt.% down to 3.2 wt.%, meanwhile for 20 wt.% of carbon almost no tin formation occurs (0.7 wt.%). Thus, we demonstrate that carbon behaves as an efficient PCA agent: it prevents reaction $R_1$ by avoiding surface contact between Si and $Ni_{3.4}Sn_4$ powders.

Beside its PCA role on milling, carbon is known to have beneficial effects on electrochemical properties of Si-based electrodes enabling to accommodate volume changes and to insure good electronic conductivity [46,47]. However, it provides low electrochemical capacity as compared to silicon or tin. Therefore, carbon content should be optimized. Moreover, in the present investigation, it is observed that the amount of carbon is decisive to the formation of a homogeneous matrix embedding Si-nanoparticles. As mentioned earlier, for a constant Si-amount, this homogeneous matrix is efficiently formed at low C-contents (9 and 13 wt.%) but not at high C-ones (17 and 20 wt.%). Interestingly, the capacity fading in the C-series shows an inverse correlation with C-amount and therefore with the formation of the homogeneous matrix. Capacity fading gradually increases from 0.024 at 0.073 %/cycle when C-content increases from 9 to 20 wt.% (Table 3). This result empirically demonstrates that the homogeneous $Ni_{3.4}Sn_4$-C matrix efficiently accommodates Si volume changes on cycling ensuring long cycle-life in composites Si-$Ni_{3.4}Sn_4$-Al-C.

Beside carbon as milling PCA agent, conductive binder and matrix former, silicon plays a key electrochemical role as it provides most of the capacity for the composite. Formally, increasing Si content should increase the capacity. This is indeed the case for the first lithiation of 10 and 20 wt.% of Si (at constant carbon amount of 13 wt.%) that provides 794 and 1191 $mAhg^{-1}$, respectively. However, increasing the Si amount to 30 wt.% does not bring further capacity and the first discharge capacity, 1186 $mAhg^{-1}$, remains far from the upper-limit capacity 1535 $mAhg^{-1}$. In fact, the 30wt.%-Si composite exhibits experimental capacities very similar to those of 20 wt.% Si (Table 3, Fig. 8). Several factors may account for this result. First, as the Si content increases from 10 to 30 wt.%, the amount of $Ni_{3.4}Sn_4$ decreases from 74 to 54 wt.%, reducing the efficiency of the latter phase to accommodate Si volume expansion. Second, when Si-content increases, the extent of reaction $R_1$ increases on milling and more $NiSi_2$ is formed (up to 6 wt.% for composite 30 wt.% Si). This partially consumes Si to produce electrochemically inactive $NiSi_2$.

To summarize, the highest electrochemical capacity with good cycle-life performances has been found for composite materials formed by Si-nanoparticles embedded in a homogenous $Ni_{3.4}Sn_4$-C matrix. This corresponds to C-contents in the range 9-13wt.% and Si-contents in the range 20-30 wt.%. Nanocomposites have reversible capacity that well exceeds that of graphite and allow to sustain reasonable cycle life for hundreds of cycles (Fig. 6 and Fig. 8).

## 5. Conclusion

Seven composites have been prepared by milling four components (Si, $Ni_{3.4}Sn_4$, Al and C) varying the amounts of silicon (10 to 30 wt.%) and carbon (0 to 20 wt.%). All composites contain Si nanoparticles surrounded by a nanostructured matrix made of $Ni_{3.4}Sn_4$, Sn, $NiSi_2$, disordered carbon and aluminum. This matrix is chemically homogeneous at the submicrometric level, and mainly consist of nanometric $Ni_{3.4}Sn_4$ and carbon, for C-contents in the range 9-13wt.% and Si-contents in the range 20-30 wt.%. The electrochemical profiles show that the reaction mechanism occurs in two main steps, i) the reaction of the intermetallic with Li between 0.6 and 0.3 V to form $Li_ySn$ phases and ii) the reaction of Li with silicon between 0.25 and 0.09 V to form $Li_ySi$ phases. The formation of $Li_ySn$ phases only occurs when the composite matrix contains nanometric $Ni_{3.4}Sn_4$. During milling, carbon addition prevents formation of Sn and $NiSi_2$ by impeding the reaction occurring between Si and $Ni_{3.4}Sn_4$. In addition, carbon in combination with $Ni_{3.4}Sn_4$ bring electronic conductivity and volume change accommodation of Si upon (de-)lithiation. However, carbon content should be limited to the range 9-13wt.% as higher content is detrimental to form a homogeneous $Ni_{3.4}Sn_4$/C matrix that effectively accommodates Si volume changes during electrochemical cycling. As concerns silicon, higher amount brings better capacity, but a saturation is observed at 20 wt.%-Si with increased capacity fading upon cycling attributed to loss of strain accommodation and conductivity enhancement on reducing the amount of $Ni_{3.4}Sn_4$/C matrix. An optimum Si/matrix ratio should be preserved to overcome these issues. Furthermore, this work supports previous studies suggesting that $Ni_{3.4}Sn_4$ needs to be highly nanostructured to provide significant reversible capacity and long-cycle-life [15,17,48], and demonstrates that, when used as a buffering matrix for Si-type anodes, it must form a homogenous matrix surrounding Si nanoparticles. Optimization of the electrode formulation and testing in full cell configuration are still needed to fully develop this anode materials for practical Li-ion batteries.

Acknowledgment

The authors are grateful to Remy PIRES for SEM and EDX analysis and to the French Research Agency ANR (project NEWMASTE, n°ANR-13-PRGE-0010) for financial support.

**Table and figure captions**

**Table 1.** Composition (in wt.%) of Si-Ni-Sn-Al-C composites milled 20 hours with different carbon and silicon contents

**Table 2.** Crystallographic data for nanostructured composites. Ni over-stoichiometry (x) in Ni3+xSn4 and crystallite size (L) for all phases are given. Standard deviations refereed to the last digit are given in parenthesis.

**Table 3.** Electrochemical properties for the composites milled 20 hours with different carbon and silicon contents. $C_{upper-limit}$ stands for the maximum expected capacity, $C_{lith,1st}$ and $C_{delith,1st}$ are the lithiation and delithiation capacities at first cycle, $C_{irrev,1st}$ is the irreversible capacity at first cycle and $C_{rev,3rd}$ and $C_{rev,1000th}$ are the reversible capacities at cycles 3 and 1000.

**Fig. 1.** XRD patterns of the carbon-free composite before (bottom) and after (top) 20h of milling. Position of diffraction lines for Sn, Si, $NiSi_2$ and $Ni_{3.4}Sn_4$ phases as reported in Pearson's crystal data base [36] are shown in the bottom part of the figure.

**Fig. 2.** XRD patterns for 20 wt.% Si composites with C-content varying from 9 wt.% (bottom) to 20wt.% (top).

**Fig. 3.** XRD patterns for 13 wt.% C composites and Si-content varying from 10 wt.% (bottom) to 30 wt.% (top).

**Fig. 4.** SEM-BSE cross section micrographs of the carbon-containing composites milled for 20 hours.

**Fig. 5.** SEM-BSE image (left) of the 9 wt.% C - 20 wt.% Si composite showing three different compositional areas. EDX point analyses of the three areas are displayed on the right-hand side.

**Fig. 6.** Reversible capacity on cycling for the four composites at constant Si-amount (20 wt.%) and increasing carbon content from 9 to 20 wt.%. Only reference cycles are shown. The first point corresponds to cycle 3 considered as the first reference cycle after two initial activation cycles.

**Fig. 7.** Profiles of the first galvanostatic cycles (a) and dQ/dV dependence as a function of voltage for the third cycle (b) for the 9 wt.% and 20wt. % carbon-containing composites. (Si-content = 20 wt.%).

**Fig. 8.** Reversible capacity on cycling for the composites with constant C-amount (13wt.%) and increasing Si-content from 10 to 30 wt.%. Only reference cycles are shown. The first point corresponds to cycle 3 considered as the first reference cycle after two initial activation cycles.

**Fig. 9.** Profiles of the first galvanostatic cycles (a) and *dQ/dV* dependence as a function of voltage for the third cycle (b) for the 10 wt.% and 30 wt.% silicon-containing composites (C-content = 13 wt.%).

# Table 1

| | 10 wt.% Si | 20 wt.% Si | 30 wt.% Si |
|---|---|---|---|
| 0 wt.% C | | $Si_{0.22}Ni_{0.22}Sn_{0.53}Al_{0.03}$ | |
| 9 wt.% C | | $Si_{0,20}Ni_{0,20}Sn_{0,48}Al_{0,03}C_{0,09}$ | |
| 13 wt.% C | $Si_{0,10}Ni_{0,22}Sn_{0,52}Al_{0,03}C_{0,13}$ | $Si_{0,19}Ni_{0,19}Sn_{0,46}Al_{0,03}C_{0,13}$ | $Si_{0,30}Ni_{0,16}Sn_{0,38}Al_{0,03}C_{0,13}$ |
| 17 wt.% C | | $Si_{0,18}Ni_{0,18}Sn_{0,44}Al_{0,03}C_{0,17}$ | |
| 20 wt.% C | | $Si_{0,18}Ni_{0,18}Sn_{0,42}Al_{0,03}C_{0,20}$ | |

# Table 2

| Composition C (wt%) | Composition Si (wt%) | Phase | Content (wt.%) | S.G. | Cell parameters $a$(Å) | Cell parameters $b$(Å) | Cell parameters $c$(Å) | β (°) | $x$ values $Ni_{3+x}Sn_4$ | $L$ (nm) |
|---|---|---|---|---|---|---|---|---|---|---|
| 0 | 20 | $Ni_{3+x}Sn_4$ | 9 (1) | *C*2/*m* | 12.199* | 4.0609* | 5.2238* | 105.17* | 0.1* | 10* |
| | | Si | 13 (1) | *Fd*-3*m* | 5.430* | | | | | 15 (2) |
| | | Sn | 43 (1) | *I*4₁/*amd* | 5.8303 (2) | | 3.1822 (1) | | | 27 (1) |
| | | $NiSi_2$ | 35 (2) | *Fm*-3*m* | 5.4731 (5) | | | | | 5 (1) |
| 9 | 20 | $Ni_{3+x}Sn_4$ | 79(1) | *C*2/*m* | 12.273 (1) | 4.0421 (4) | 5.2007 (5) | 104.73 (1) | 0.29 (3) | 8 (1) |
| | | Si | 14 (1) | *Fd*-3*m* | 5.430* | | | | | 14 (2) |
| | | Sn | 3.2 (2) | *I*4₁/*amd* | 5.830* | | 3.182* | | | 22* |
| | | $NiSi_2$ | 4 (1) | *Fm*-3*m* | 5.470* | | | | | 6* |
| 13 | 20 | $Ni_{3+x}Sn_4$ | 75 (1) | *C*2/*m* | 12.299 (1) | 4.0497 (3) | 5.2043 (4) | 104.67 (1) | 0.29 (3) | 8 (1) |
| | | Si | 17 (1) | *Fd*-3*m* | 5.430* | | | | | 14 (1) |
| | | Sn | 2.5 (4) | *I*4₁/*amd* | 5.830* | | 3.182* | | | 22* |
| | | $NiSi_2$ | 5 (1) | *Fm*-3*m* | 5.470* | | | | | 6* |
| 17 | 20 | $Ni_{3+x}Sn_4$ | 74 (1) | *C*2/*m* | 12.410 (1) | 4.0685 (3) | 5.2010 (6) | 104.04 (1) | 0.48 (3) | 8 (1) |
| | | Si | 22 (1) | *Fd*-3*m* | 5.430* | | | | | 14 (1) |
| | | Sn | 0.9 (2) | *I*4₁/*amd* | 5.830* | | 3.182* | | | 22* |
| | | $NiSi_2$ | 2 (1) | *Fm*-3*m* | 5.470* | | | | | 6* |
| 20 | 20 | $Ni_{3+x}Sn_4$ | 72 (1) | *C*2/*m* | 12.452 (1) | 4.0800 (1) | 5.2089 (2) | 103.60 (1) | 0.60 (3) | 50 (3) |
| | | Si | 27 (1) | *Fd*-3*m* | 5.430* | | | | | 17 (2) |
| | | Sn | 0.7 (2) | *I*4₁/*amd* | 5.830* | | 3.182* | | | 22* |
| | | $NiSi_2$ | 0 (1) | *Fm*-3*m* | 5.470* | | | | | 6* |
| 13 | 10 | $Ni_{3+x}Sn_4$ | 90 (2) | *C*2/*m* | 12.392 (2) | 4.0679 (4) | 5.1994 (6) | 104.18 (1) | 0.47 (3) | 9 (1) |
| | | Si | 9 (1) | *Fd*-3*m* | 5.430* | | | | | 14 (3) |
| | | Sn | 0.5 (2) | *I*4₁/*amd* | 5.830* | | 3.182* | | | 22* |
| | | $NiSi_2$ | 0 (1) | *Fm*-3*m* | 5.470* | | | | | 6* |
| 13 | 30 | $Ni_{3+x}Sn_4$ | 57 (2) | *C*2/*m* | 12.287(2) | 4.0466(6) | 5.2079(8) | 104.65(1) | 0.37 (5) | 7 (1) |
| | | Si | 31 (1) | *Fd*-3*m* | 5.430* | | | | | 11 (2) |
| | | Sn | 6 (0.3) | *I*4₁/*amd* | 5.830* | | 3.182* | | | 22* |
| | | $NiSi_2$ | 6 (1) | *Fm*-3*m* | 5.470* | | | | | 6* |

*Values were fixed to ensure refinement stability due to strong peak overlapping (Si and $NiSi_2$ phases) or low phase amount

**Table 3**

| C/Si (wt.%) | $C_{upper-limit}$ (mAh g$^{-1}$) | $C_{lith,1st}$ (mAhg$^{-1}$) | $C_{lith,1st}$ / $C_{upper-limit}$ (%) | $C_{delith,1st}$ (mAhg$^{-1}$) | $C_{irrev, 1st}$ (%) | $C_{rev, 3rd}$ (mAhg$^{-1}$) | $C_{rev, 1000th}$ (mAhg$^{-1}$) | Capacity fade (%/cycle)* |
|---|---|---|---|---|---|---|---|---|
| 0/20 | 1348 | 949 | 70 | 716 | 25 | 510 | 213[a] | 0.28[a] |
| 9/20 | 1260 | 1196 | 95 | 985 | 18 | 707 | 542 | 0.024 |
| 13/20 | 1219 | 1191 | 98 | 1008 | 15 | 698 | 429 | 0.039 |
| 17/20 | 1178 | 1146 | 97 | 956 | 17 | 711 | 311 | 0.056 |
| 20/20 | 1170 | 923 | 79 | 702 | 24 | 522 | 136 | 0.073 |
| 13/10 | 955 | 794 | 83 | 572 | 28 | 388 | 256[b] | 0.077[b] |
| 13/30 | 1535 | 1186 | 77 | 966 | 17 | 700 | 487[c] | 0.082[c] |

*Fading values are calculated from cycle 3 to cycle 1000 except for *a*, *b* and *c* compositions for which last measured cycle was 206, 450 and 372, respectively.

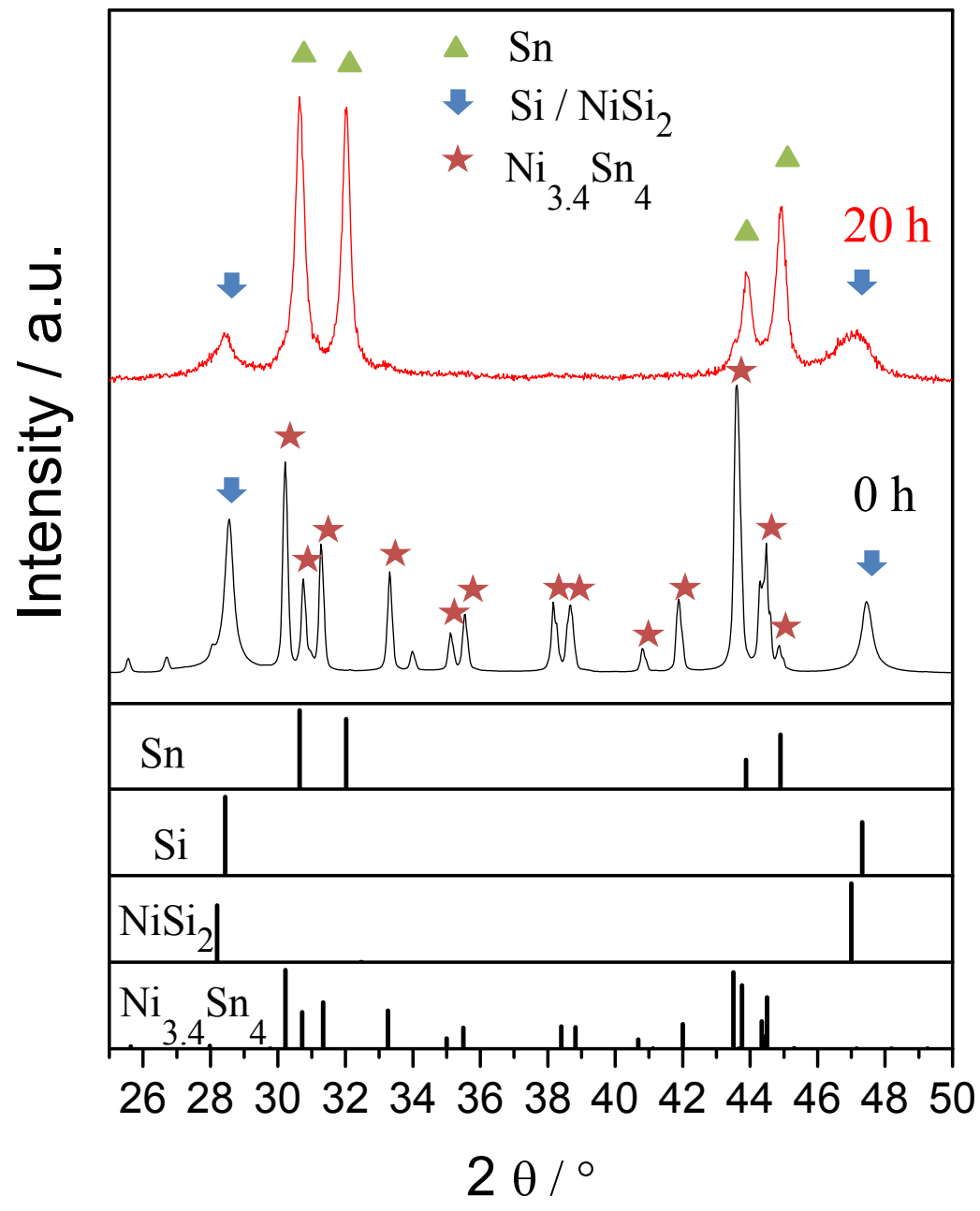

**Fig. 1.**

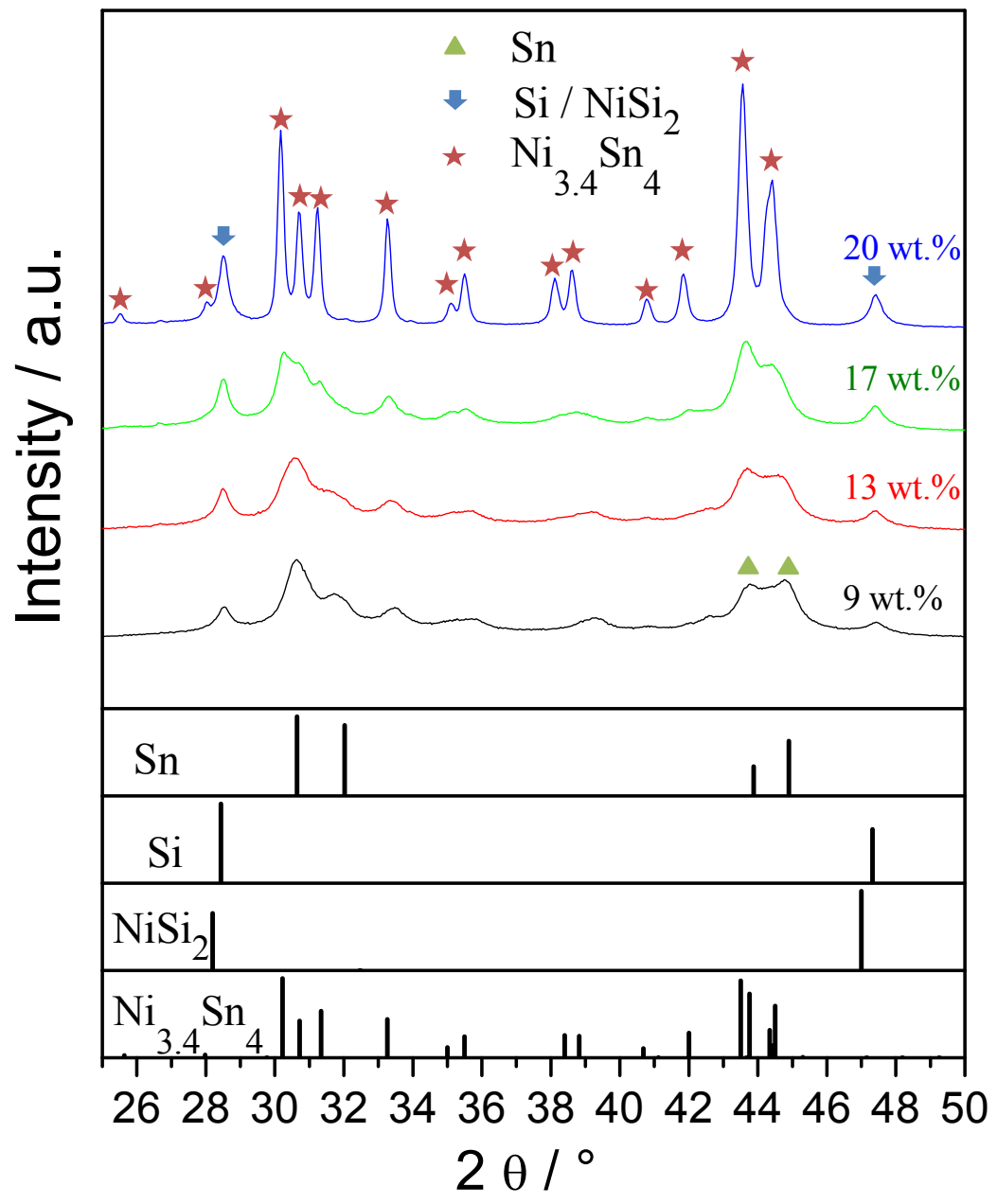
Sn
Si / NiSi2
Ni3.4Sn4
20 wt.%
17 wt.%
13 wt.%
9 wt.%
Intensity / a.u.
Sn
Si
NiSi2
Ni3.4Sn4
26 28 30 32 34 36 38 40 42 44 46 48 50
2 θ / °

**Fig. 2.**

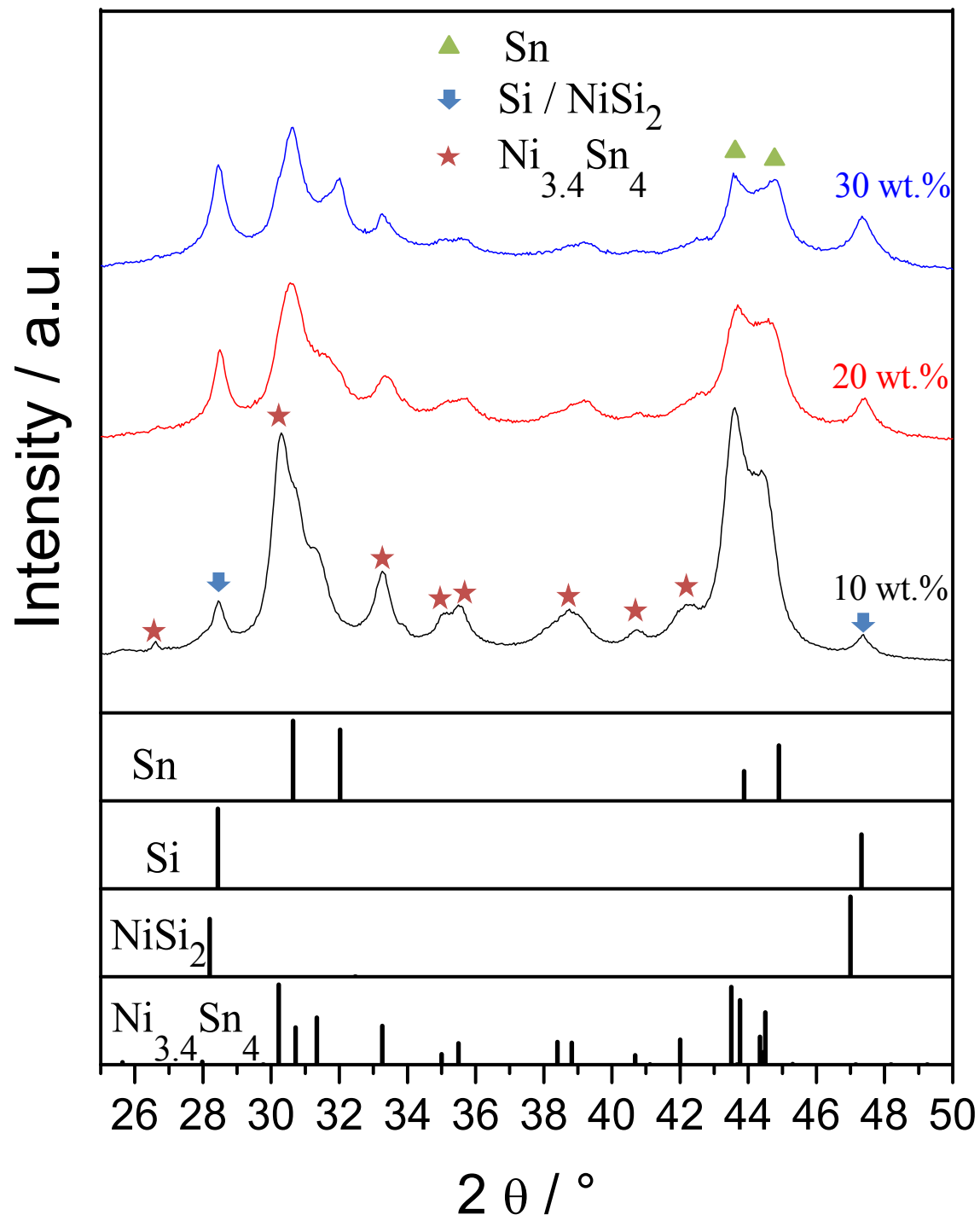

Sn
Si / NiSi2
Ni3.4Sn4
30 wt.%
20 wt.%
10 wt.%
Intensity / a.u.
Sn
Si
NiSi2
Ni3.4Sn4
26 28 30 32 34 36 38 40 42 44 46 48 50
2 θ / °

**Fig. 3.**

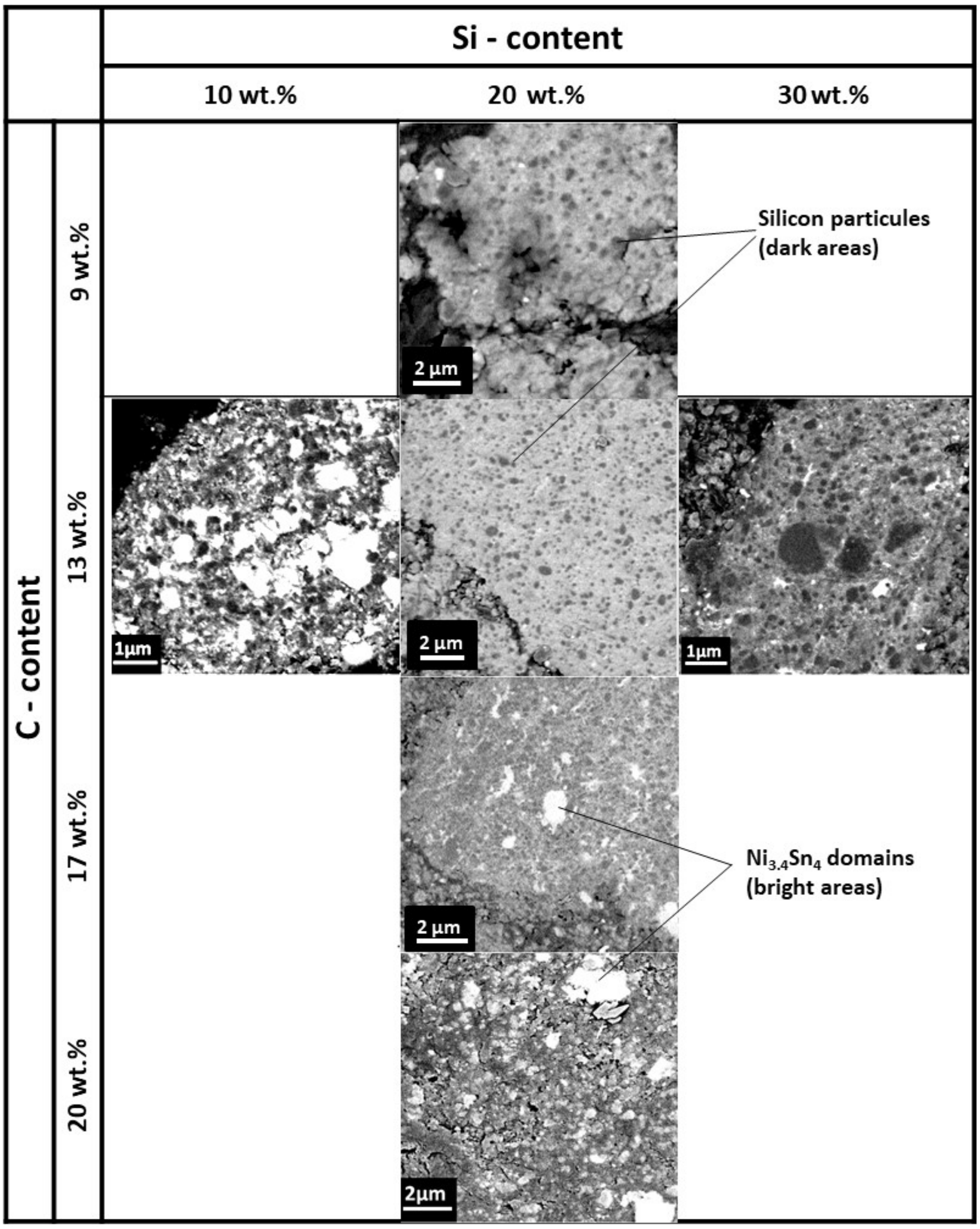

**Fig. 4.**

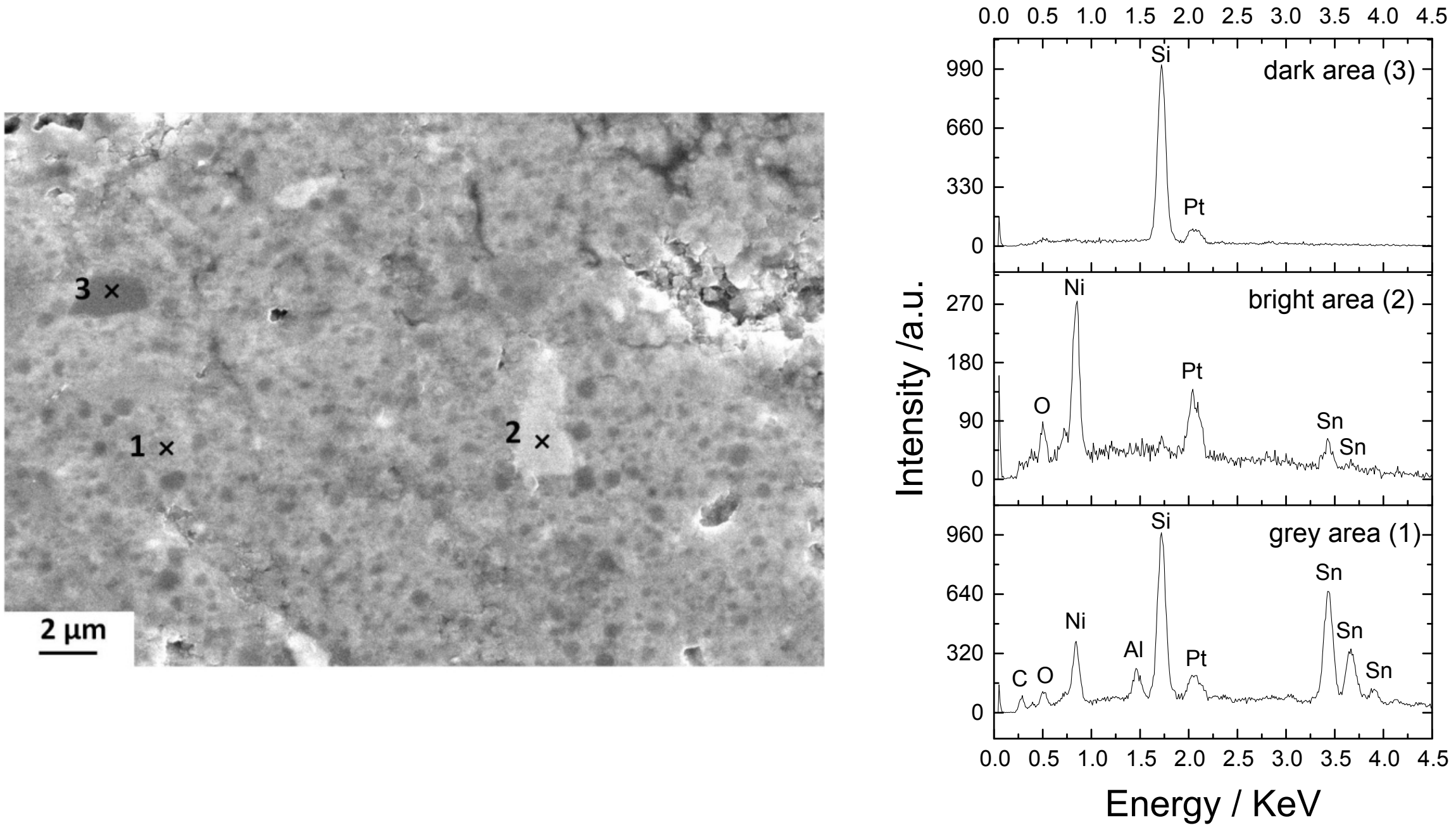
3 ×
1 ×
2 ×
2 µm
0.0 0.5 1.0 1.5 2.0 2.5 3.0 3.5 4.0 4.5
dark area (3)
Si
Pt
bright area (2)
Ni
O
Pt
Sn
Sn
grey area (1)
Si
Sn
Ni
Al
Pt
C O
Sn
Sn
Intensity /a.u.
Energy / KeV

**Fig. 5.**

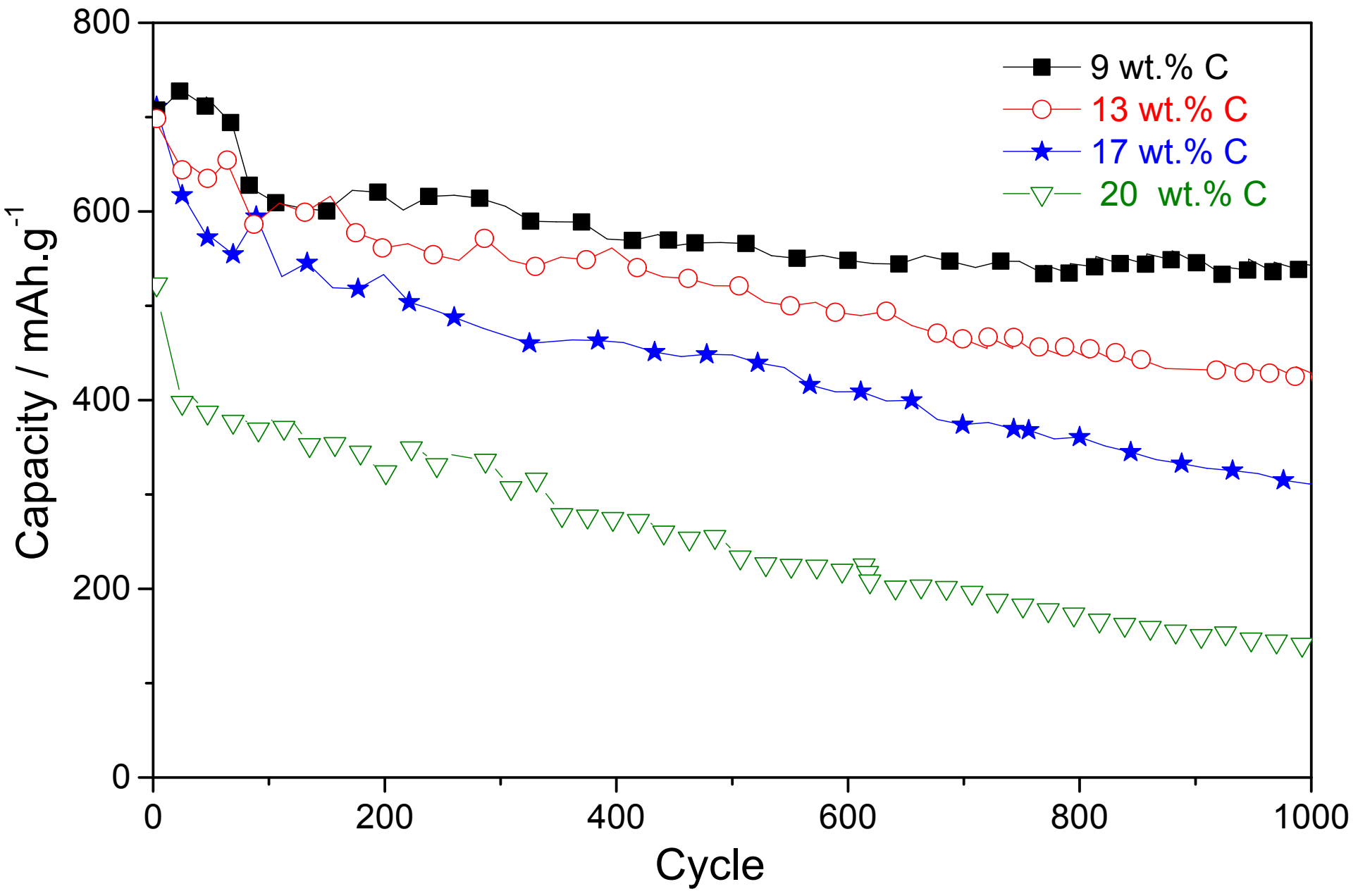

**Fig. 6.**

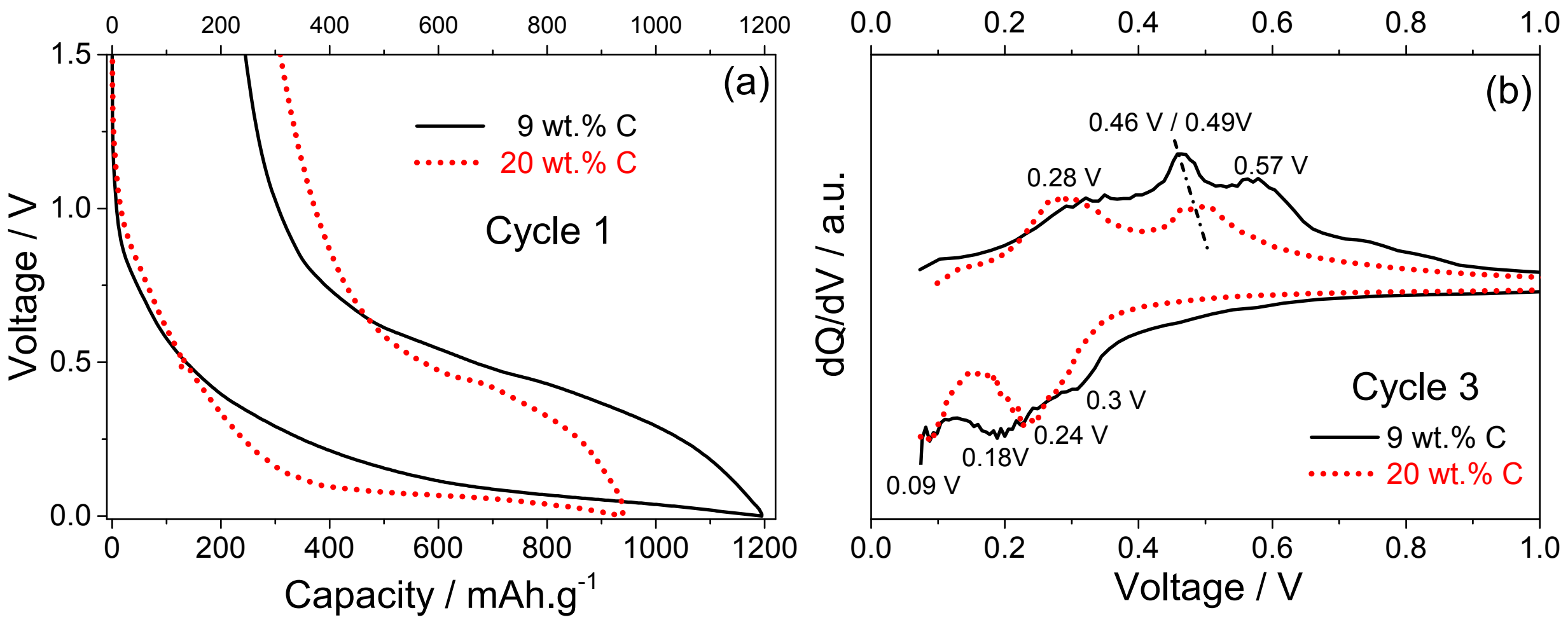

**Fig. 7.**

**Fig. 8.**

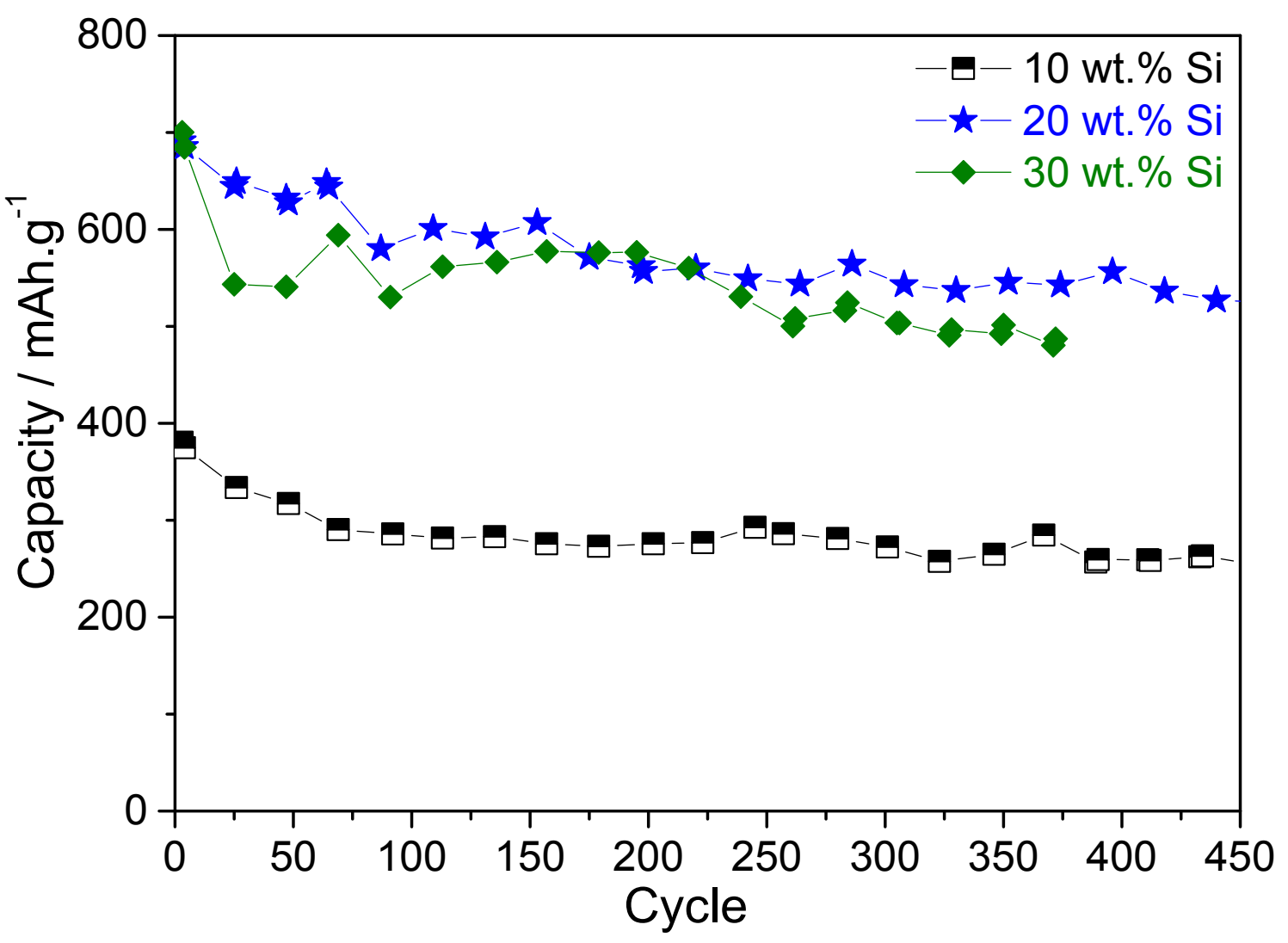
10 wt.% Si
20 wt.% Si
30 wt.% Si
Capacity / mAh.g-1
Cycle

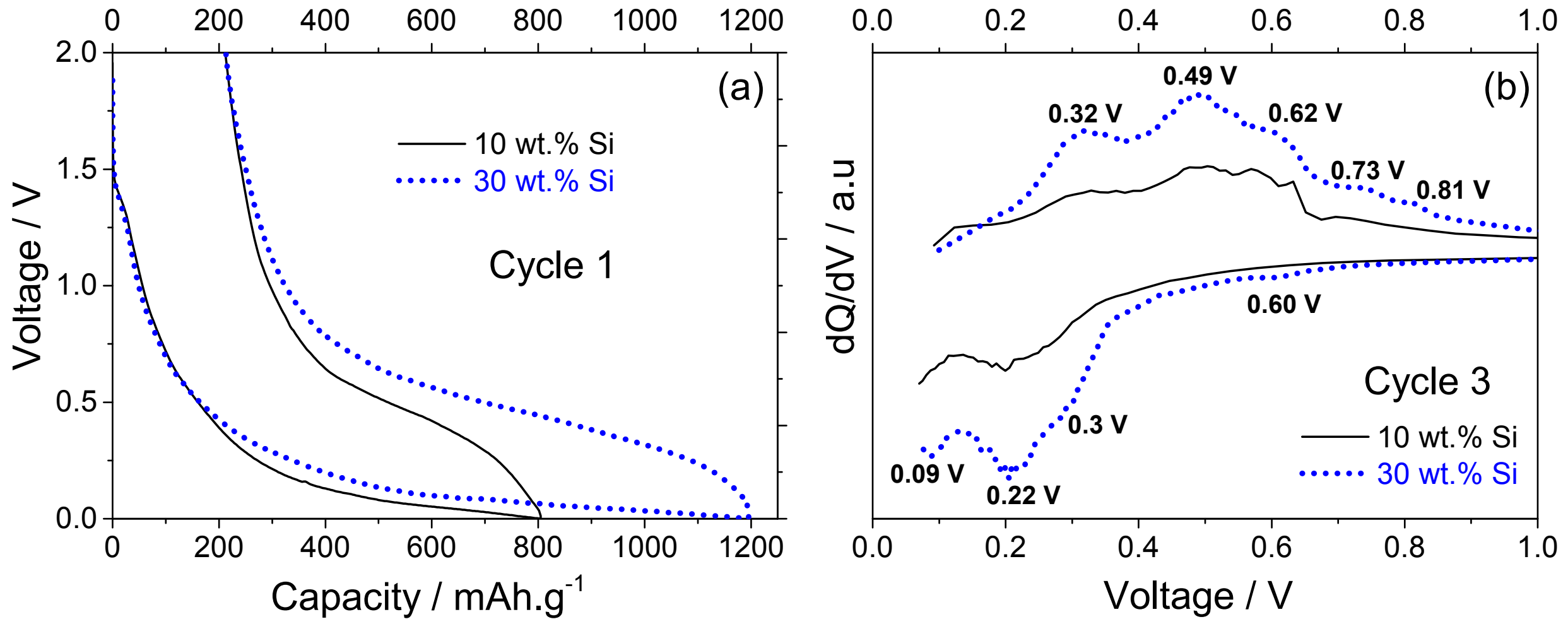

**Fig. 9.**

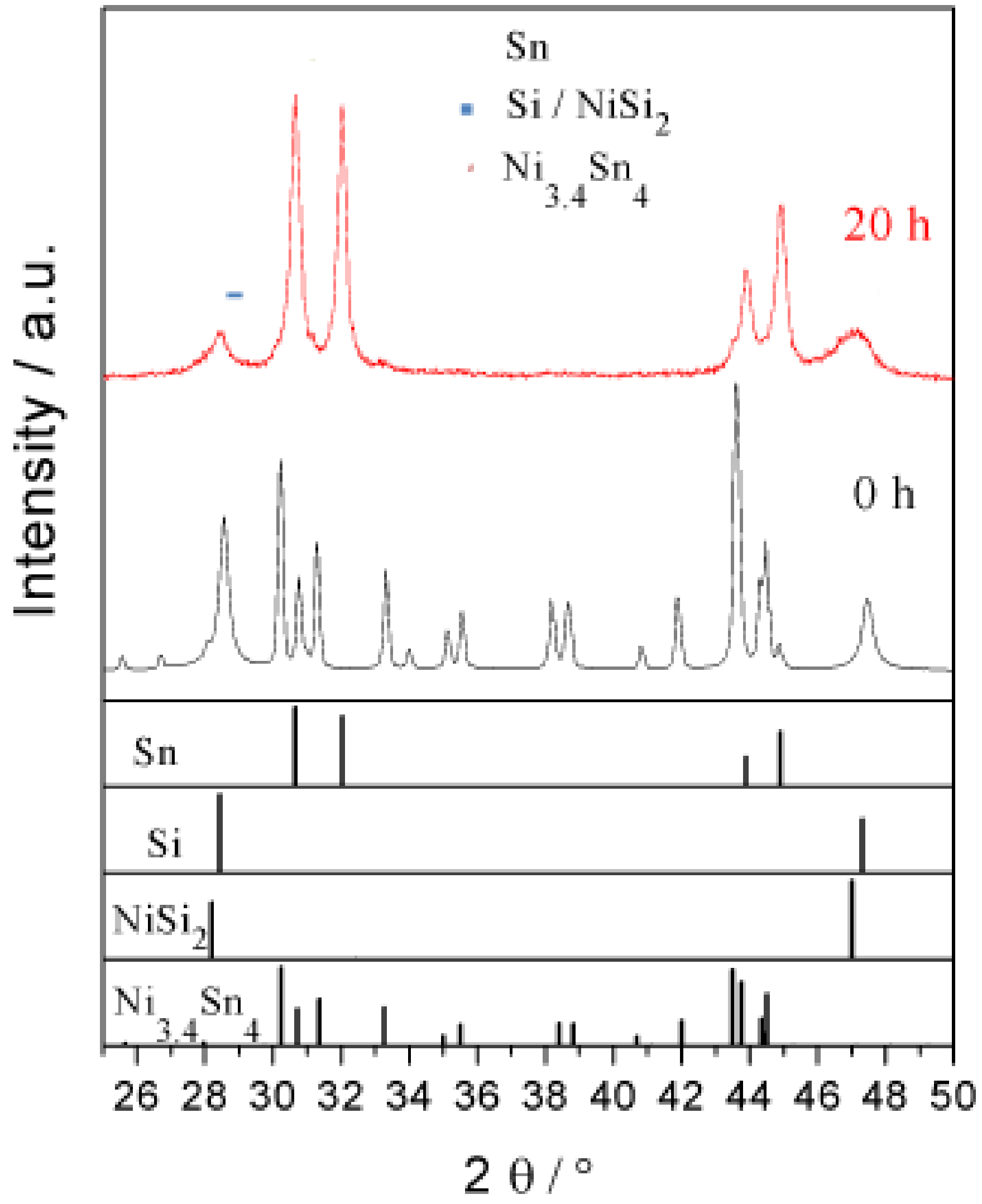

**Fig. 1.** XRD patterns of the carbon-free composite before (bottom) and after (top) 20h of milling. Position of diffraction lines for Sn, Si, $NiSi_2$ and $Ni_{3.4}Sn_4$ phases as reported in Pearson's crystal data base (Villar and Cenzual, 2010) are shown in the bottom part of the figure.

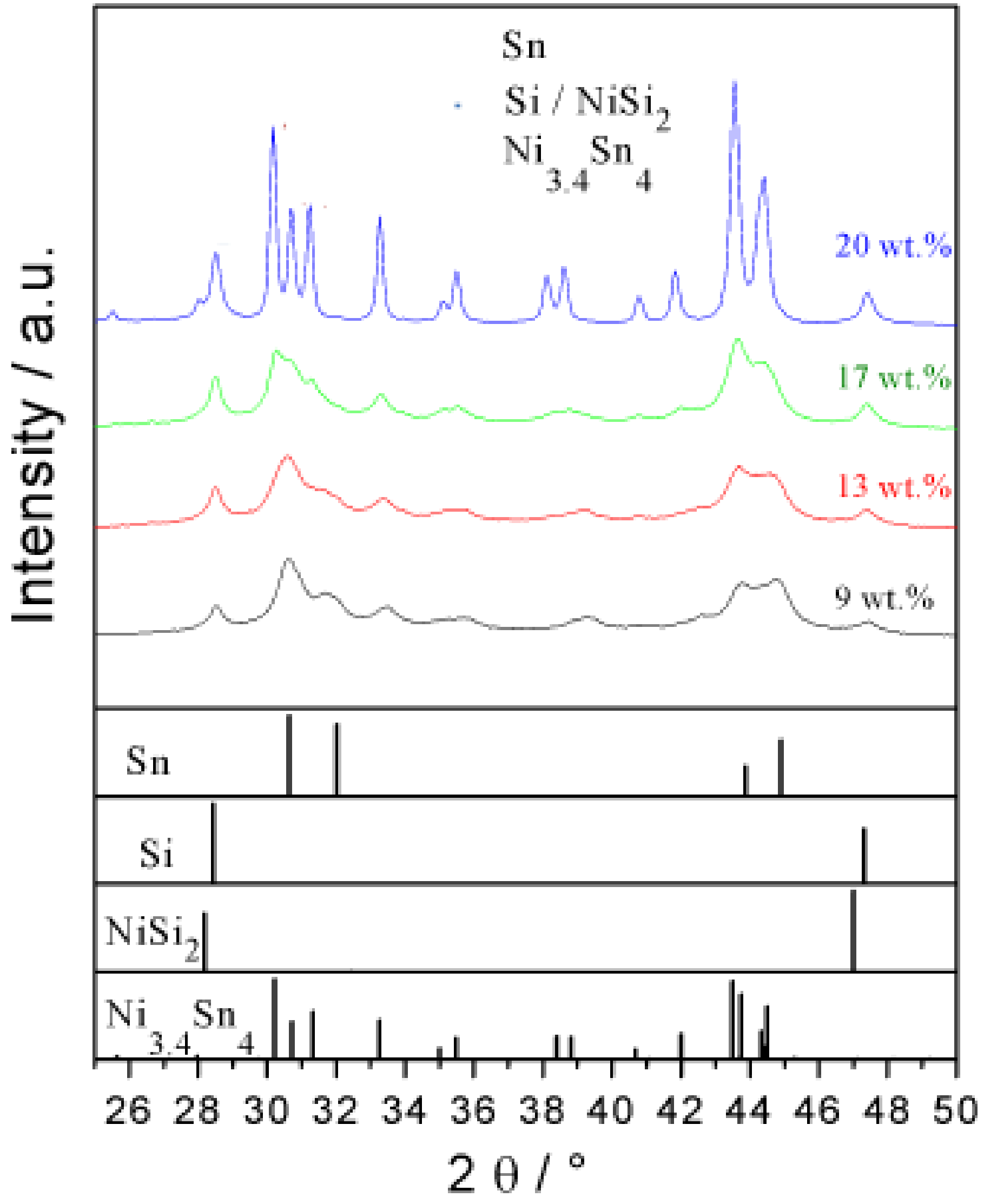

**Fig. 2.** XRD patterns for20 wt.% Si composites with C-content varying from 9 wt.% (bottom) to 20wt.% (top).

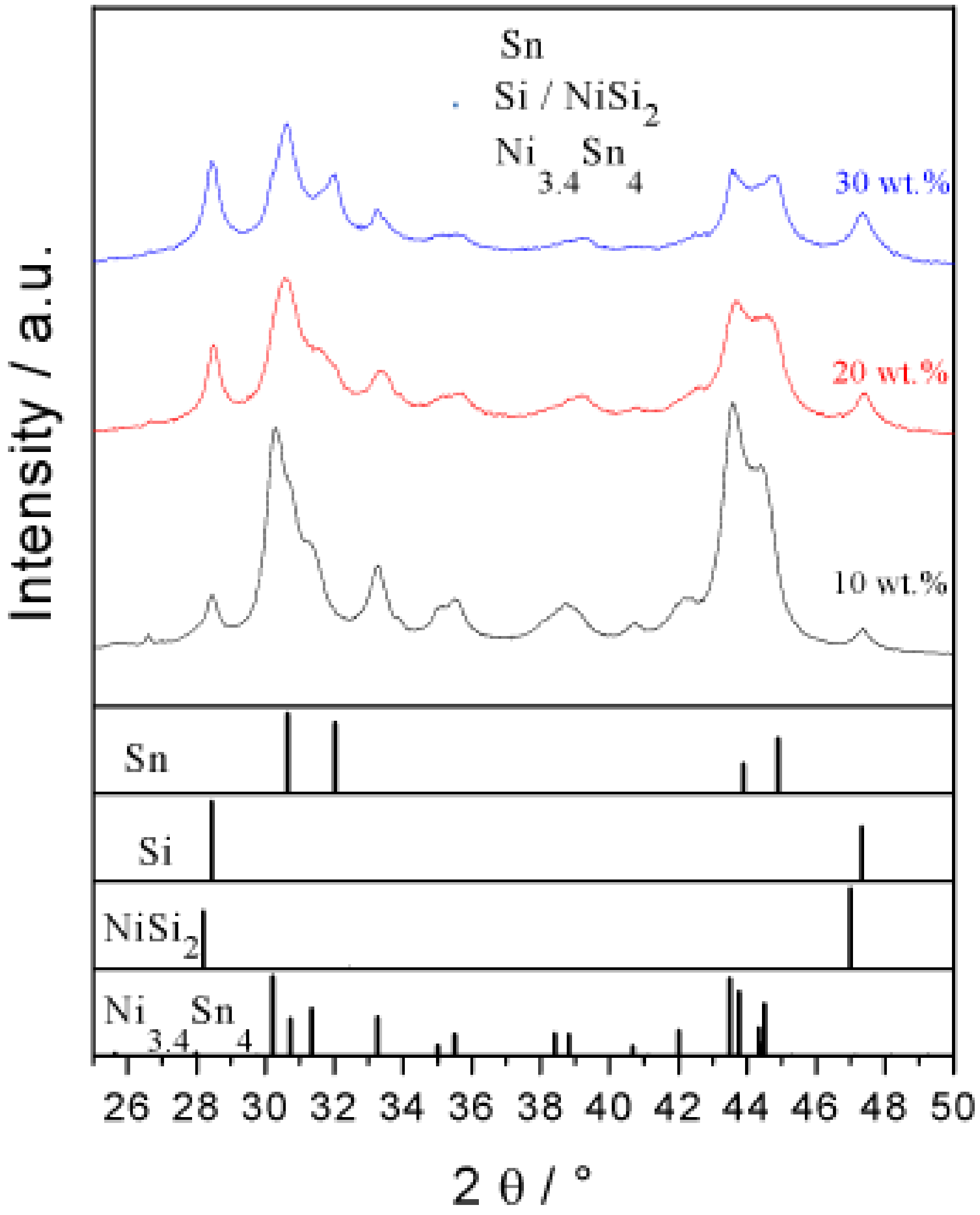

**Fig. 3.** XRD patterns for 13 wt.% C composites and Si-content varying from 10 wt.% (bottom) to 30 wt.% (top).

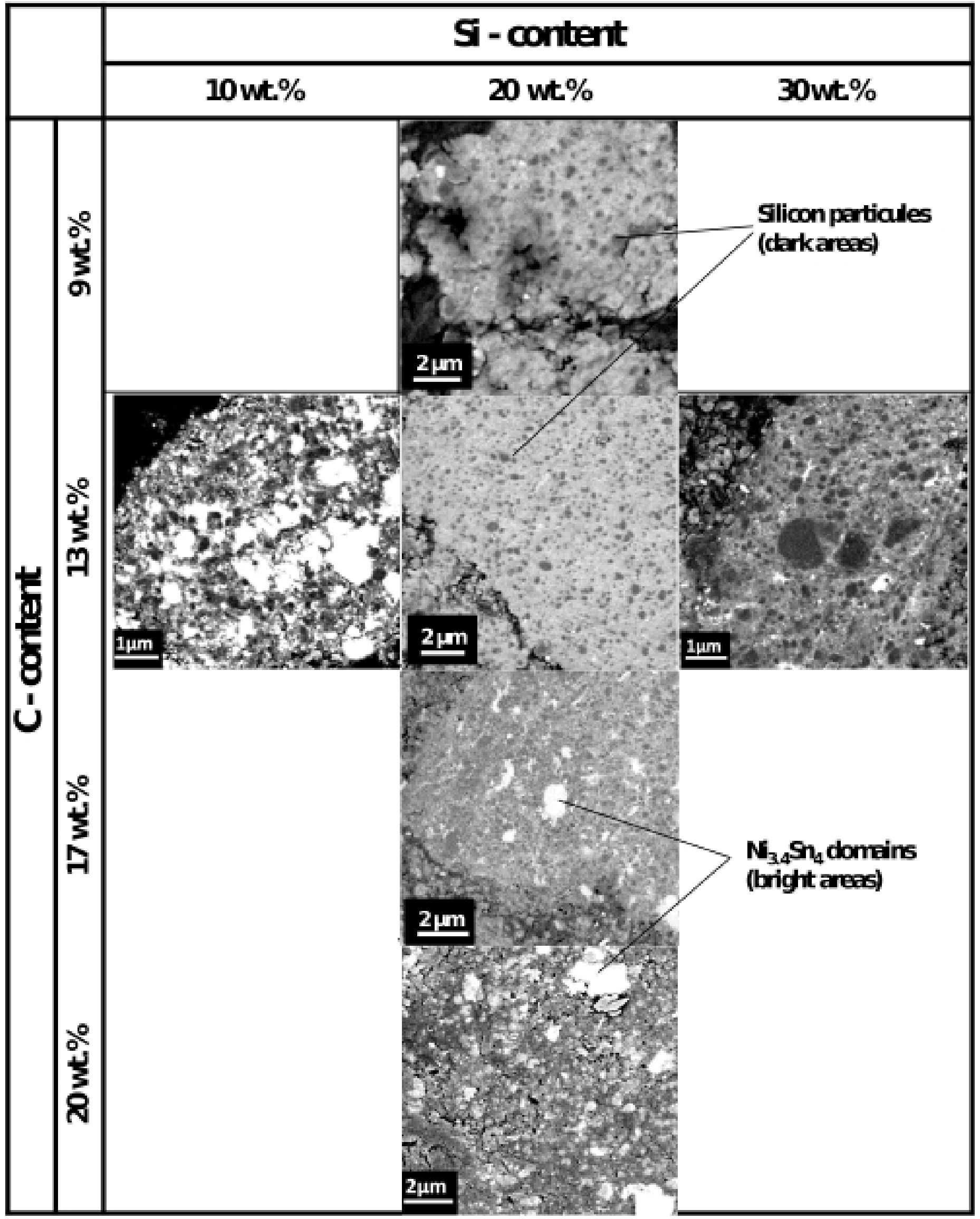

**Fig. 4.** SEM-BSE cross section micrographs of the carbon-containing composites milled for 20 hours.

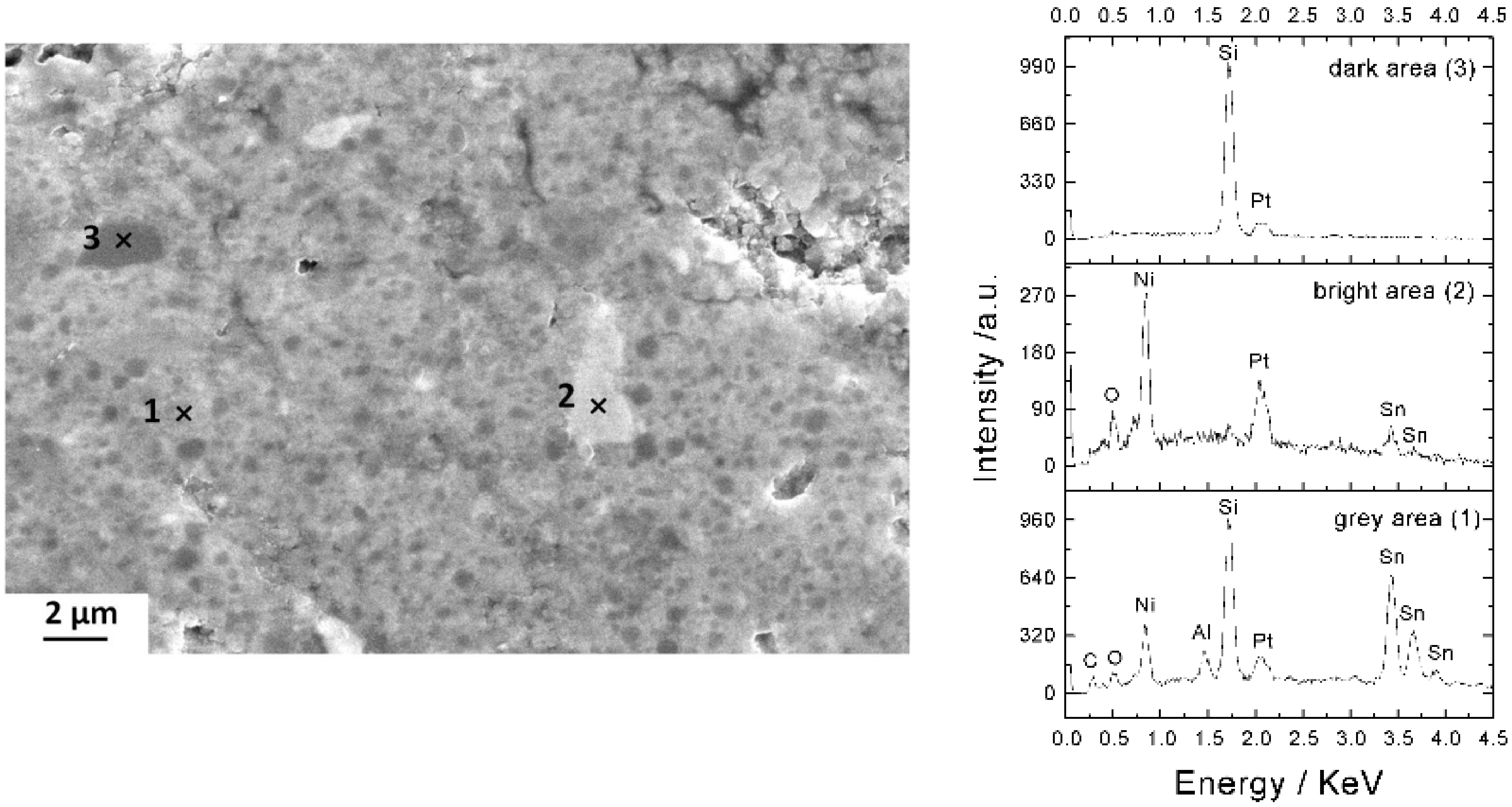

**Fig. 5.** SEM-BSE image (left) of the 9 wt.% C - 20 wt.% Si composite showing three different compositional areas. EDX point analyses of the three areas are displayed on the right-hand side.

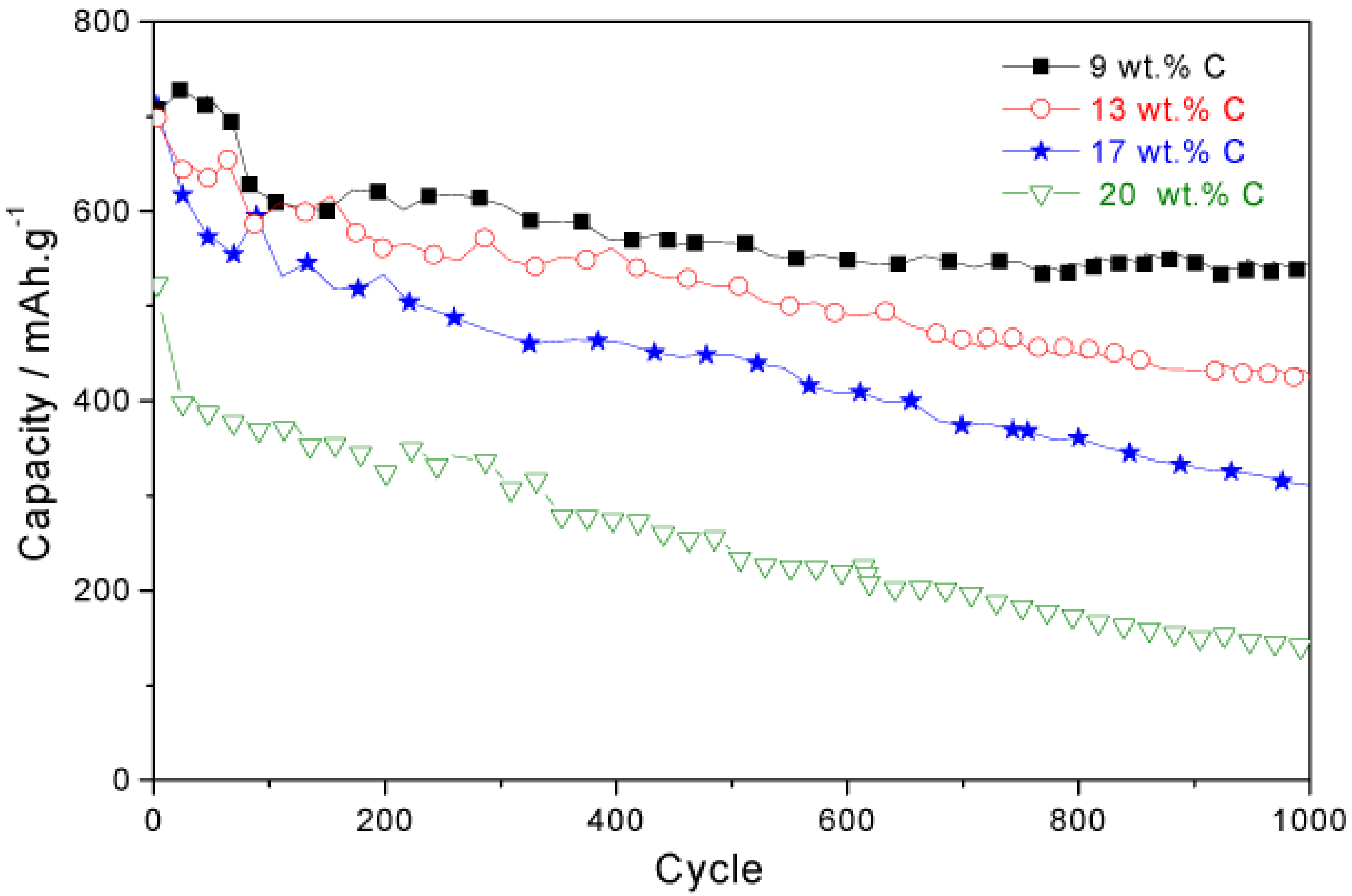

**Fig. 6.** Reversible capacity on cycling for the four composites at constant Si-amount (20 wt.%) and increasing carbon content from 9 to 20 wt.%. Only reference cycles are shown. The first point corresponds to cycle 3 considered as the first reference cycle after two initial activation cycles.

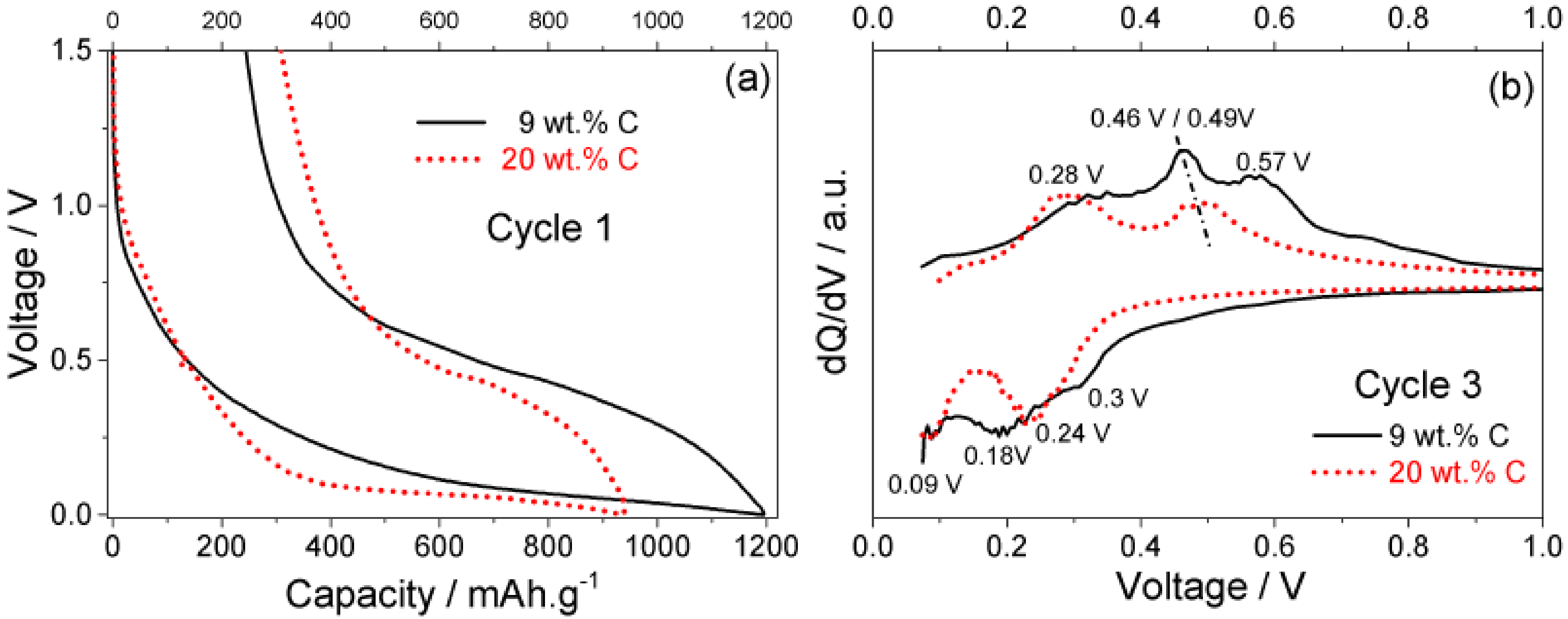

**Fig. 7.** Profiles of the first galvanostatic cycles (a) and dQ/dV dependence as a function of voltage for the third cycle (b) for the 9 wt.% and 20wt. % carbon-containing composites. (Si-content = 20 wt.%).

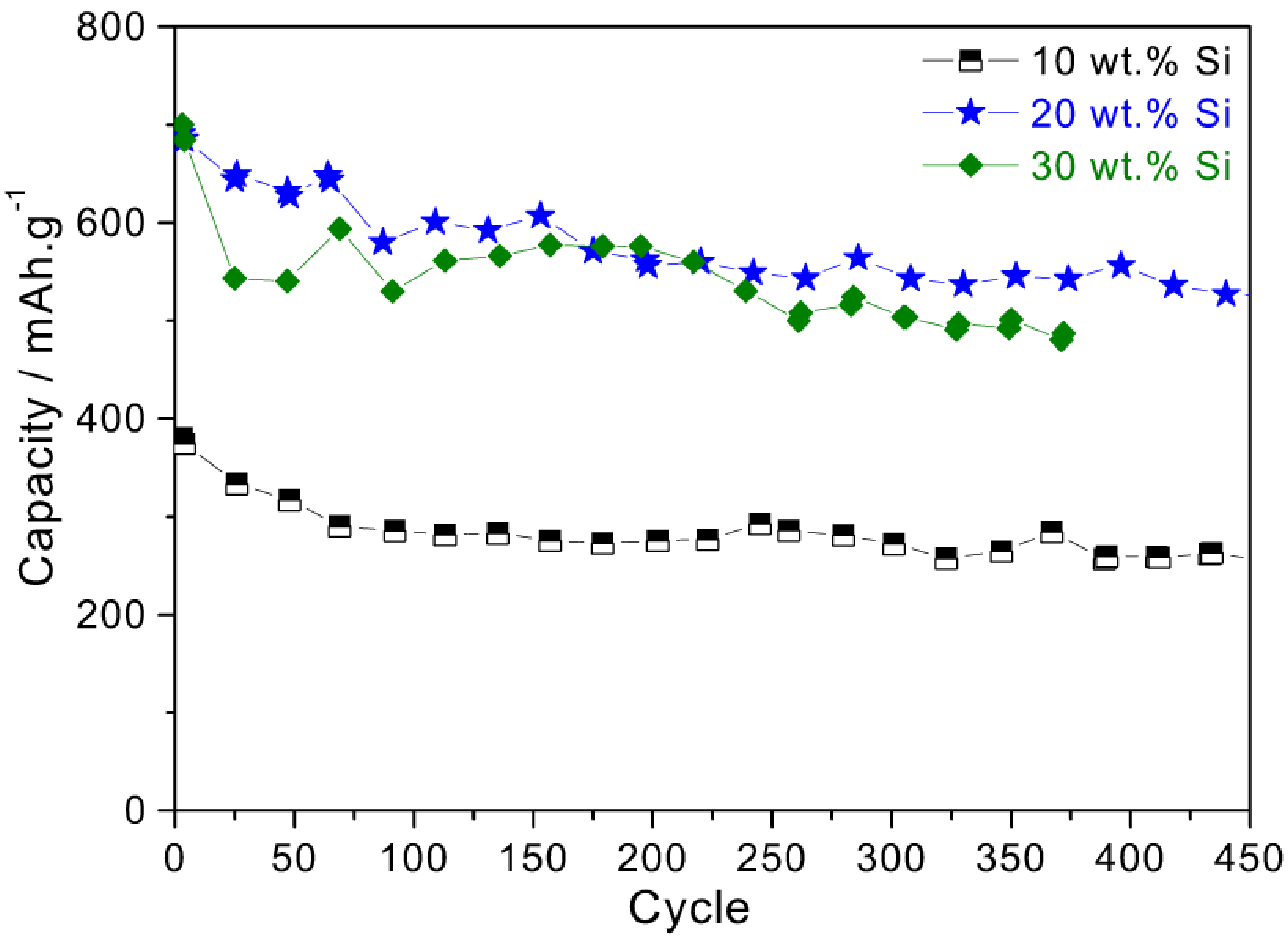

**Fig. 8.** Reversible capacity on cycling for the composites with constant C-amount (13wt.%) and increasing Si-content from 10 to 30 wt.%. Only reference cycles are shown. The first point corresponds to cycle 3 considered as the first reference cycle after two initial activation cycles.

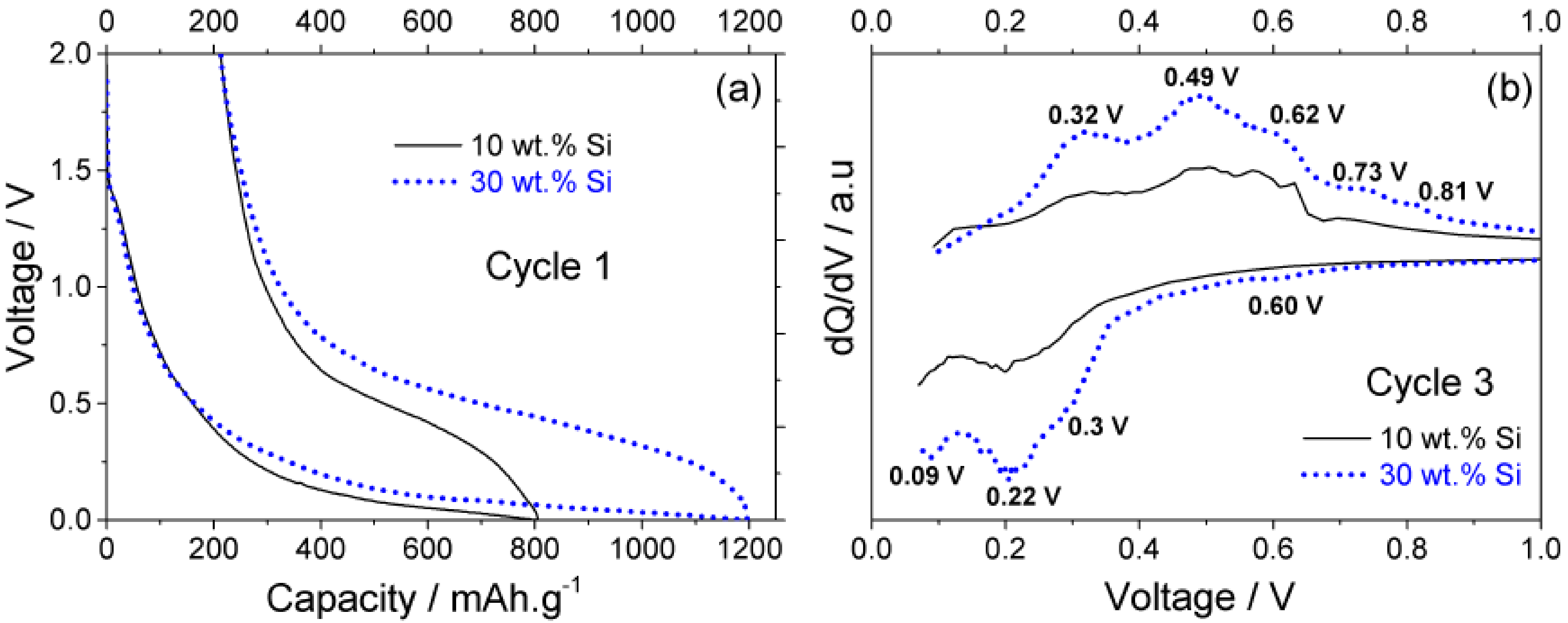

**Fig. 9.** Profiles of the first galvanostatic cycles (a) and *dQ*/*dV* dependence as a function of voltage for the third cycle (b) for the 10 wt.% and 30 wt.% silicon-containing composites (C-content = 13 wt.%).

**Table 1.** Composition (in wt.%) of Si-Ni-Sn-Al-C composites milled 20 hours with different carbon and silicon contents

| | 10 wt.% Si | 20 wt.% Si | 30 wt.% Si |
|---|---|---|---|
| 0 wt.% C | | $Si_{0.22}Ni_{0.22}Sn_{0.53}Al_{0.03}$ | |
| 9 wt.% C | | $Si_{0,20}Ni_{0,20}Sn_{0,48}Al_{0,03}C_{0,09}$ | |
| 13 wt.% C | $Si_{0,10}Ni_{0,22}Sn_{0,52}Al_{0,03}C_{0,13}$ | $Si_{0,19}Ni_{0,19}Sn_{0,46}Al_{0,03}C_{0,13}$ | $Si_{0,30}Ni_{0,16}Sn_{0,38}Al_{0,03}C_{0,13}$ |
| 17 wt.% C | | $Si_{0,18}Ni_{0,18}Sn_{0,44}Al_{0,03}C_{0,17}$ | |
| 20 wt.% C | | $Si_{0,18}Ni_{0,18}Sn_{0,42}Al_{0,03}C_{0,20}$ | |

**Table 1.** Crystallographic data for nanostructured composites. Ni over-stoichiometry ($x$) in $Ni_{3+x}Sn_4$ and crystallite size ($L$) for all phases are given. Standard deviations refereed to the last digit are given in parenthesis**.**

| Composition C (wt%) | Composition Si (wt%) | Phase | Content (wt.%) | S.G. | Cell parameters a(Å) | b(Å) | c(Å) | β (°) | x values $Ni_{3+x}Sn_4$ | L (nm) |
|---|---|---|---|---|---|---|---|---|---|---|
| 0 | 20 | $Ni_{3+x}Sn_4$ | 9 (1) | *C*2/*m* | 12.199* | 4.0609* | 5.2238* | 105.17* | 0.1* | 10* |
| | | Si | 13 (1) | *Fd*-3*m* | 5.430* | | | | | 15 (2) |
| | | Sn | 43 (1) | *I*4$_1$/*amd* | 5.8303 (2) | | 3.1822 (1) | | | 27 (1) |
| | | $NiSi_2$ | 35 (2) | *Fm*-3*m* | 5.4731 (5) | | | | | 5 (1) |
| 9 | 20 | $Ni_{3+x}Sn_4$ | 79(1) | *C*2/*m* | 12.273 (1) | 4.0421 (4) | 5.2007 (5) | 104.73 (1) | 0.29 (3) | 8 (1) |
| | | Si | 14 (1) | *Fd*-3*m* | 5.430* | | | | | 14 (2) |
| | | Sn | 3.2 (2) | *I*4$_1$/*amd* | 5.830* | | 3.182* | | | 22* |
| | | $NiSi_2$ | 4 (1) | *Fm*-3*m* | 5.470* | | | | | 6* |
| 13 | 20 | $Ni_{3+x}Sn_4$ | 75 (1) | *C*2/*m* | 12.299 (1) | 4.0497 (3) | 5.2043 (4) | 104.67 (1) | 0.29 (3) | 8 (1) |
| | | Si | 17 (1) | *Fd*-3*m* | 5.430* | | | | | 14 (1) |
| | | Sn | 2.5 (4) | *I*4$_1$/*amd* | 5.830* | | 3.182* | | | 22* |
| | | $NiSi_2$ | 5 (1) | *Fm*-3*m* | 5.470* | | | | | 6* |
| 17 | 20 | $Ni_{3+x}Sn_4$ | 74 (1) | *C*2/*m* | 12.410 (1) | 4.0685 (3) | 5.2010 (6) | 104.04 (1) | 0.48 (3) | 8 (1) |
| | | Si | 22 (1) | *Fd*-3*m* | 5.430* | | | | | 14 (1) |
| | | Sn | 0.9 (2) | *I*4$_1$/*amd* | 5.830* | | 3.182* | | | 22* |
| | | $NiSi_2$ | 2 (1) | *Fm*-3*m* | 5.470* | | | | | 6* |
| 20 | 20 | $Ni_{3+x}Sn_4$ | 72 (1) | *C*2/*m* | 12.452 (1) | 4.0800 (1) | 5.2089 (2) | 103.60 (1) | 0.60 (3) | 50 (3) |
| | | Si | 27 (1) | *Fd*-3*m* | 5.430* | | | | | 17 (2) |
| | | Sn | 0.7 (2) | *I*4$_1$/*amd* | 5.830* | | 3.182* | | | 22* |
| | | $NiSi_2$ | 0 (1) | *Fm*-3*m* | 5.470* | | | | | 6* |
| 13 | 10 | $Ni_{3+x}Sn_4$ | 90 (2) | *C*2/*m* | 12.392 (2) | 4.0679 (4) | 5.1994 (6) | 104.18 (1) | 0.47 (3) | 9 (1) |
| | | Si | 9 (1) | *Fd*-3*m* | 5.430* | | | | | 14 (3) |
| | | Sn | 0.5 (2) | *I*4$_1$/*amd* | 5.830* | | 3.182* | | | 22* |
| | | $NiSi_2$ | 0 (1) | *Fm*-3*m* | 5.470* | | | | | 6* |
| 13 | 30 | $Ni_{3+x}Sn_4$ | 57 (2) | *C*2/*m* | 12.287(2) | 4.0466(6) | 5.2079(8) | 104.65(1) | 0.37 (5) | 7 (1) |
| | | Si | 31 (1) | *Fd*-3*m* | 5.430* | | | | | 11 (2) |
| | | Sn | 6 (0.3) | *I*4$_1$/*amd* | 5.830* | | 3.182* | | | 22* |
| | | $NiSi_2$ | 6 (1) | *Fm*-3*m* | 5.470* | | | | | 6* |

*Values were fixed to ensure refinement stability due to strong peak overlapping (Si and $NiSi_2$ phases) or low phase amount

**Table 3.** Electrochemical properties for the composites milled 20 hours with different carbon and silicon contents. $C_{upper\text{-}limit}$ stands for the maximum expected capacity, $C_{lith,1st}$ and $C_{delith,1st}$ are the lithiation and delithiation capacities at first cycle, $C_{irrev,1st}$ is the irreversible capacity at first cycle and $C_{rev,3rd}$ and $C_{rev,1000th}$ are the reversible capacities at cycles 3 and 1000.

| C/Si (wt.%) | $C_{upper\text{-}limit}$ (mAh g$^{-1}$) | $C_{lith,1st}$ (mAhg$^{-1}$) | $C_{lith,1st}$ / $C_{upper\text{-}limit}$ (%) | $C_{delith,1st}$ (mAhg$^{-1}$) | $C_{irrev,1st}$ (%) | $C_{rev,3rd}$ (mAhg$^{-1}$) | $C_{rev,1000th}$ (mAhg$^{-1}$) | Capacity fade (%/cycle)* |
|---|---|---|---|---|---|---|---|---|
| 0/20 | 1348 | 949 | 70 | 716 | 25 | 510 | 213[a] | 0.28[a] |
| 9/20 | 1260 | 1196 | 95 | 985 | 18 | 707 | 542 | 0.024 |
| 13/20 | 1219 | 1191 | 98 | 1008 | 15 | 698 | 429 | 0.039 |
| 17/20 | 1178 | 1146 | 97 | 956 | 17 | 711 | 311 | 0.056 |
| 20/20 | 1170 | 923 | 79 | 702 | 24 | 522 | 136 | 0.073 |
| 13/10 | 955 | 794 | 83 | 572 | 28 | 388 | 256[b] | 0.077[b] |
| 13/30 | 1535 | 1186 | 77 | 966 | 17 | 700 | 487[c] | 0.082[c] |

*Fading values are calculated from cycle 3 to cycle 1000 except for *a*, *b* and *c* compositions for which last measured cycle was 206, 450 and 372, respectively.

**Declaration of interests**

 The authors declare that they have no known competing financial interests or personal relationships that could have appeared to influence the work reported in this paper.

☐The authors declare the following financial interests/personal relationships which may be considered as potential competing interests:

**Tahar Azib:** Conceptualization, Methodology, Validation, Writing - Original Draft **Claire Thaury:** Investigation, Conceptualization, Writing - Original Draft **Cécile Fariaut-Georges:** Investigation, Validation **Thierry Hézèque:** Resources, Data Curation **Fermin Cuevas:** Conceptualization, Methodology, Project administration, Funding acquisition, Writing - Original Draft **Christian Jordy:** Conceptualization, Methodology, Funding acquisition, **Michel Latroche:** Conceptualization, Methodology, Project administration, Funding acquisition, Writing - Original Draft

# SUPPLEMENTARY INFORMATIONS

## Role of silicon and carbon on the structural and electrochemical properties of Si-$Ni_{3.4}Sn_4$-Al-C anodes for Li-ion batteries

Tahar Aziba[a*], Claire Thaury[a,b*], Cécile Fariaut-Georges[a], Thierry Hézèque[b,] Fermin Cuevas[a], Christian Jordy[b] and Michel Latroche[a]

[a]Univ. Paris Est Creteil, CNRS, ICMPE, UMR7182, F-94320, Thiais, France.

[b]SAFT Batteries, 113 Bd. Alfred Daney, 33074 Bordeaux, France.

Figure S1 displays the graphical output of the Rietveld analysis of nanostructured composites Si-$Ni_{3.4}Sn_4$-Al-C with various carbon and silicon contents after 20-hour milling. All patterns can be indexed with main contribution of the two initials reactants (Si and $Ni_{3.4}Sn_4$), and minor contribution of two novel phases Sn and $NiSi_2$ produced during mechanical milling. Crystallographic data results of the Rietveld analysis are given in Table 2.

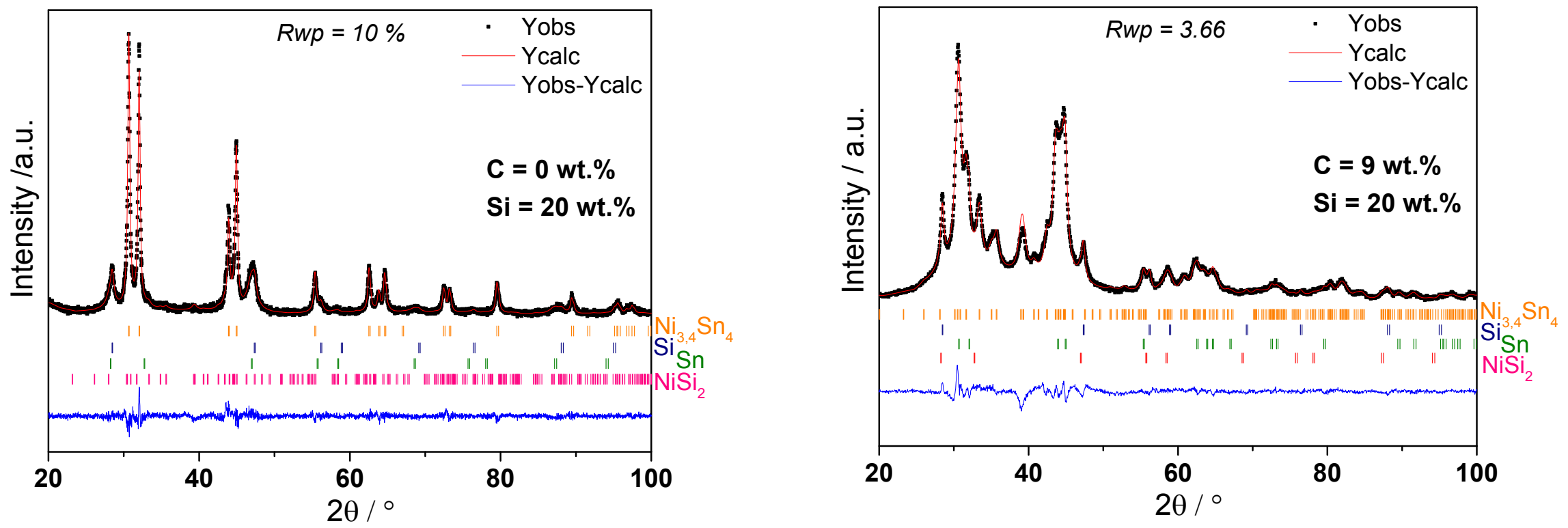

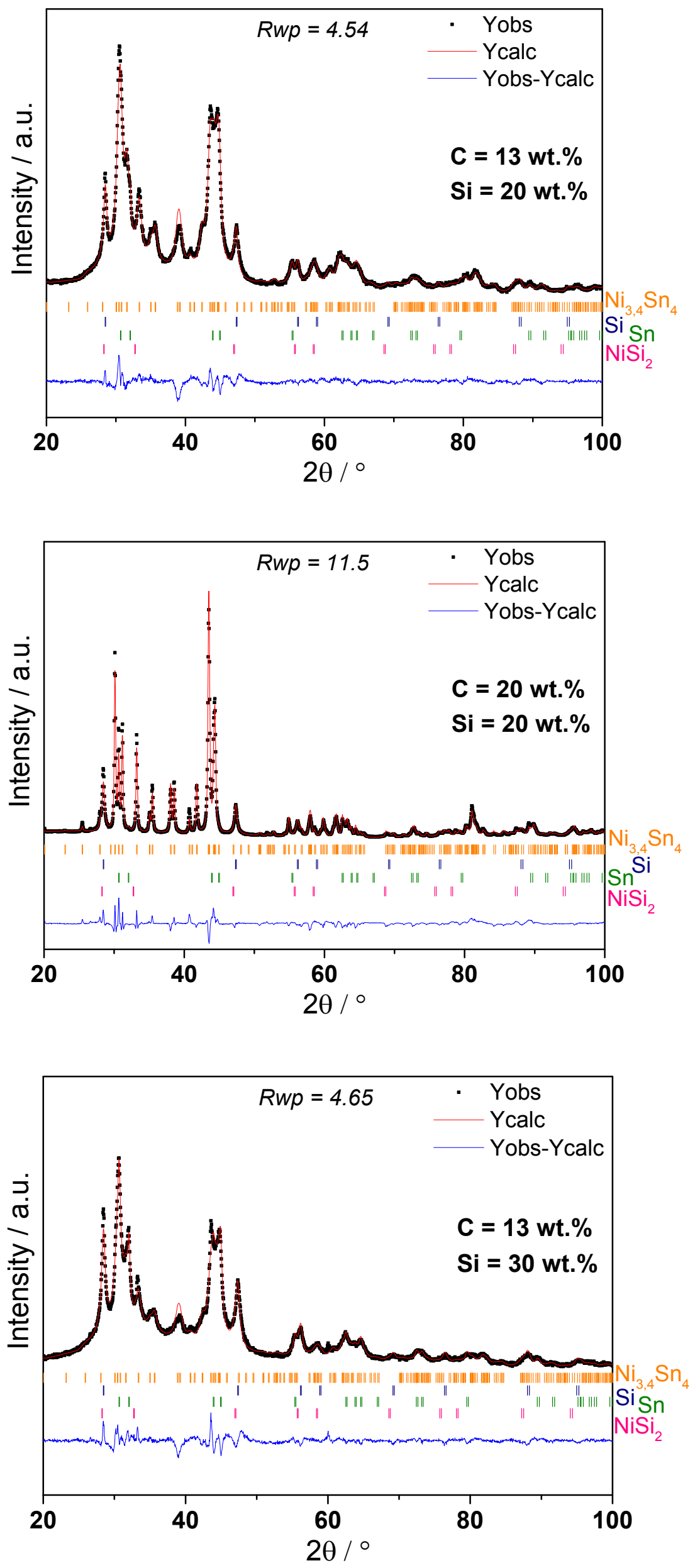

**Fig. S1.** Graphical output of Rietveld analysis for the seven composites studied in this work with various carbon and silicon contents.

Figure S2 displays the evolution of the lithiation/delithiation capacity on cycling for the carbon free composite. Rapid loss of capacity during the first cycles and poor cycling performance after 200 cycles (200 mAh/g) is observed.

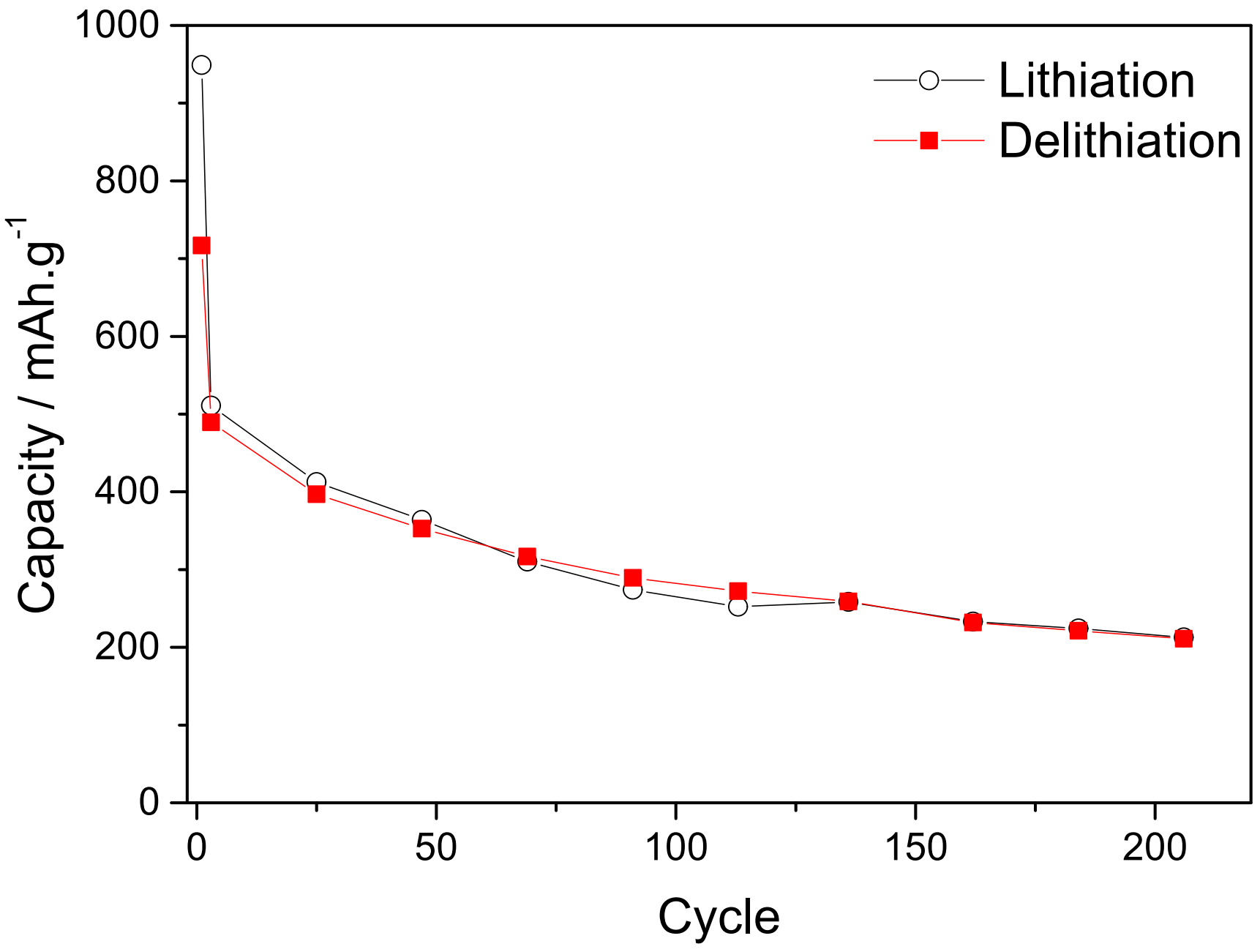

**Fig. S2.** Galvanostatic cycling for the carbon free composite. Only the first cycle and the reference ones are shown.

Figure 3 shows the coulombic efficiency for the four composites at constant Si-amount (20 wt.%) and different carbon content during 1000 cycles. The coulombic efficiency is high (~99%) and stable for all composites, though for 20 wt.% C some fluctuations are observed.

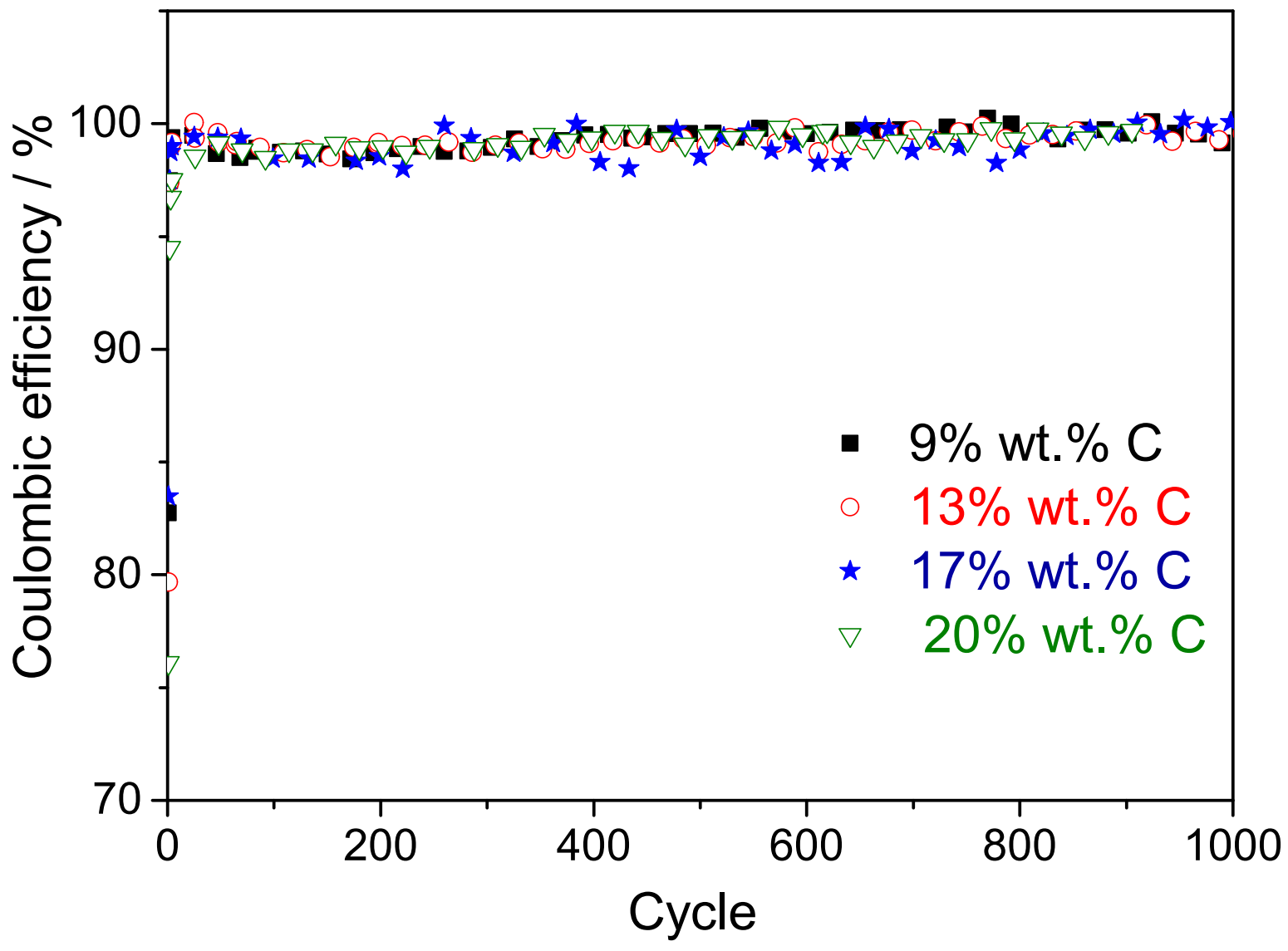

**Fig. S3** Coulombic efficiency values for the four composites at constant Si-amount (20 wt.%) and increasing carbon content from 9 to 20 wt.%.

Figure 4 displays the coulombic efficiency for the four composites at constant C-amount (20 wt.%) and different silicon content during 450 cycles. The coulombic efficiency is high (~99%) and stable for all composites. Corresponding capacities are given in Figure 8.

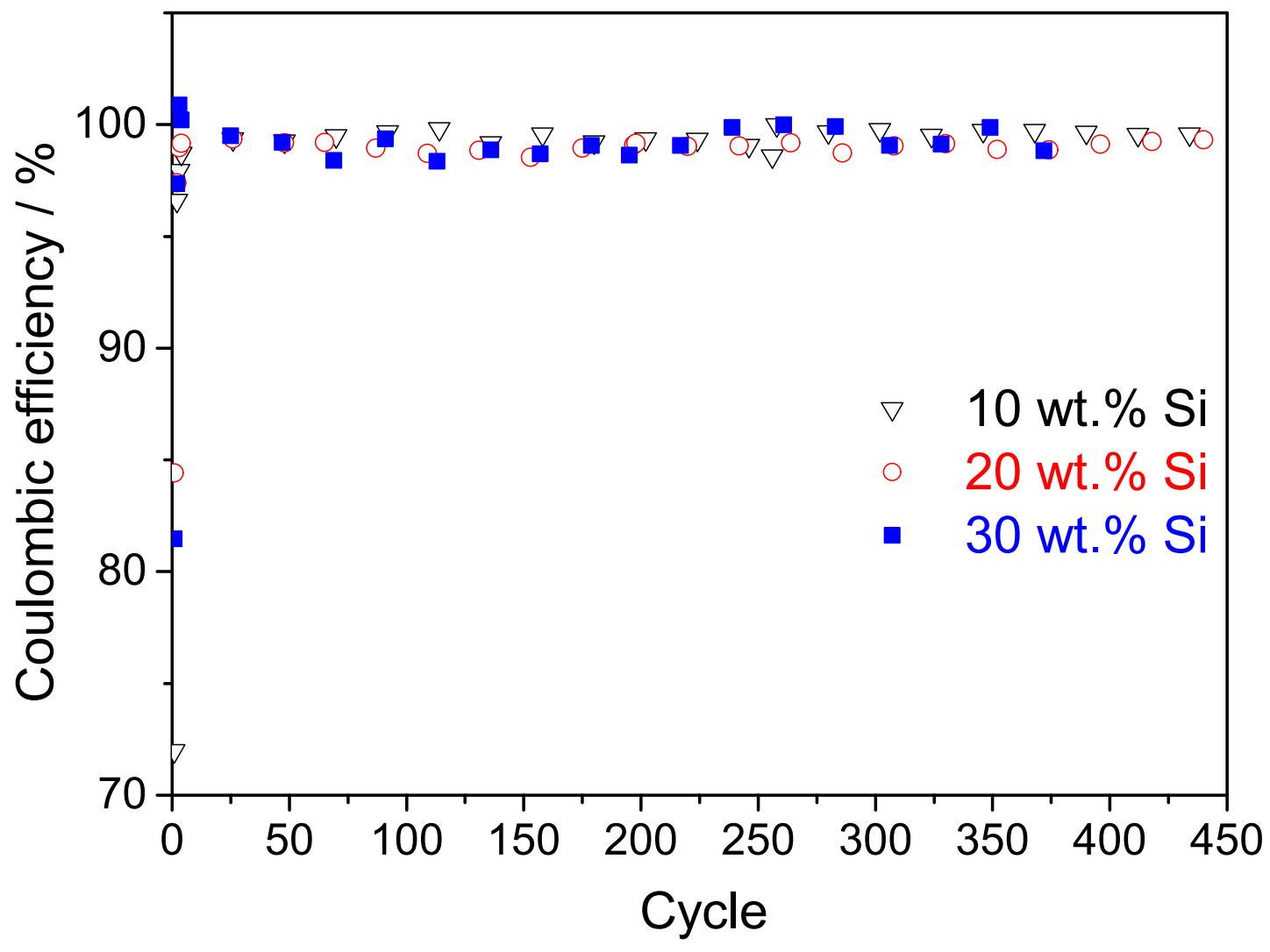

**Fig. S4.** Coulombic efficiency values for the 10 wt.% 20. wt% and 30 wt. % silicon-containing composites milled for 20 hours.